\RequirePackage[l2tabu, orthodox]{nag}
\documentclass[preprint,12pt,authoryear,3p]{elsarticle}

\usepackage{graphicx}
\usepackage{amsmath}
\usepackage{amsfonts}
\usepackage{enumitem}
\usepackage{fancyvrb}
\usepackage{color}
\usepackage{titlesec}
\usepackage{booktabs}
\usepackage{float}
\usepackage{todonotes}
\usepackage{setspace}
\usepackage{fixltx2e}
\usepackage{siunitx}
\usepackage{empheq}
\usepackage[labelfont=bf]{caption}
\usepackage{epstopdf}
\usepackage{empheq}
\usepackage{natbib}
\usepackage{natbibspacing}
\usepackage{etoolbox}
\usepackage[utf8]{inputenc}
\usepackage{datetime}
\usepackage{multirow}

\usepackage{lmodern}
\usepackage[T1]{fontenc}
\usepackage{microtype}

\usepackage[pdftex,bookmarks]{hyperref}

\usepackage{boxedminipage}

\hypersetup{colorlinks=true, linkcolor=blue, citecolor=blue, urlcolor=blue}

\makeatletter

\pretocmd{\NAT@citex}{%
  \let\NAT@hyper@\NAT@hyper@citex
  \def\NAT@postnote{#2}%
  \setcounter{NAT@total@cites}{0}%
  \setcounter{NAT@count@cites}{0}%
  \forcsvlist{\stepcounter{NAT@total@cites}\@gobble}{#3}}{}{}
\newcounter{NAT@total@cites}
\newcounter{NAT@count@cites}
\def\NAT@postnote{}

\def\NAT@hyper@citex#1{%
  \stepcounter{NAT@count@cites}%
  \hyper@natlinkstart{\@citeb\@extra@b@citeb}#1%
  \ifnumequal{\value{NAT@count@cites}}{\value{NAT@total@cites}}
    {\ifNAT@swa\else\if*\NAT@postnote*\else%
     \NAT@cmt\NAT@postnote\global\def\NAT@postnote{}\fi\fi}{}%
  \ifNAT@swa\else\if\relax\NAT@date\relax
  \else\NAT@@close\global\let\NAT@nm\@empty\fi\fi
  \hyper@natlinkend}
\renewcommand\hyper@natlinkbreak[2]{#1}

\patchcmd{\NAT@citex}
  {\ifNAT@swa\else\if*#2*\else\NAT@cmt#2\fi
   \if\relax\NAT@date\relax\else\NAT@@close\fi\fi}{}{}{}
\patchcmd{\NAT@citex}
  {\if\relax\NAT@date\relax\NAT@def@citea\else\NAT@def@citea@close\fi}
  {\if\relax\NAT@date\relax\NAT@def@citea\else\NAT@def@citea@space\fi}{}{}

\makeatother

\newcommand{\tsp}{\textsuperscript{+}}
\newcommand{\rlh}{\rightleftharpoons}
\newcommand{\Ln}{\mathbf{Ln}}

\begin{document}
\begin{frontmatter}
\title{A thermodynamic framework for modelling membrane transporters}
\author[sbl]{Michael Pan}
\ead{panm@student.unimelb.edu.au}

\author[sbl]{Peter J. Gawthrop}
\ead{peter.gawthrop@unimelb.edu.au}

\author[abi]{Kenneth Tran}
\ead{k.tran@auckland.ac.nz}

\author[wehi,mdhs]{Joseph Cursons}
\ead{cursons.j@wehi.edu.au}

\author[sbl,med,cbns]{Edmund~J.~Crampin\corref{cor1}}
\ead{edmund.crampin@unimelb.edu.au}
\address[sbl]{Systems Biology Laboratory, School of Mathematics and Statistics, and Department of Biomedical Engineering, Melbourne School of Engineering, University of Melbourne, Parkville, Victoria 3010, Australia}
\address[abi]{Auckland Bioengineering Institute, University of Auckland}
\address[wehi]{Bioinformatics Division, Walter and Eliza Hall Institute of Medical Research, Parkville, Victoria 3052, Australia}
\address[mdhs]{Department of Medical Biology, School of Medicine, University of Melbourne, Parkville, Victoria 3010, Australia}
\address[med]{School of Medicine, Faculty of Medicine, Dentistry and Health Sciences, University of Melbourne, Parkville, Victoria 3010}
\address[cbns]{ARC Centre of Excellence in Convergent Bio-Nano Science and Technology, Melbourne School of Engineering, University of Melbourne, Parkville, Victoria 3010, Australia}
\cortext[cor1]{Corresponding author}
\begin{abstract}
Membrane transporters contribute to the regulation of the internal environment of cells by translocating substrates across cell membranes. Like all physical systems, the behaviour of membrane transporters is constrained by the laws of thermodynamics. However, many mathematical models of transporters, especially those incorporated into whole-cell models, are not thermodynamically consistent, leading to unrealistic behaviour. In this paper we use a physics-based modelling framework, in which the transfer of energy is explicitly accounted for, to develop thermodynamically consistent models of transporters. We then apply this methodology to model two specific transporters: the cardiac sarcoplasmic/endoplasmic Ca\textsuperscript{2+} ATPase (SERCA) and the cardiac Na\tsp{}/K\tsp{} ATPase.
\end{abstract}

\begin{keyword}
	bond graph, biochemistry, chemical reaction network, biomedical engineering, systems biology
\end{keyword}
\end{frontmatter}

\section{Introduction}
The cell membrane is a physical barrier that separates the internal environment of a cell from its external environment, as well as many compartments within the cell. By transporting chemical species across the membrane, the cell regulates concentration within each compartment, providing the environment necessary for many cellular processes \citep{keener_mathematical_2009}. A key contributor to these transport systems are membrane transporters, which are proteins located in cell membranes that maintain cell volume \citep{glitsch_electrophysiology_2001}, set up ionic gradients required for electrical \citep{glitsch_electrophysiology_2001} and calcium signalling \citep{periasamy_serca_2007,blaustein_sodium/calcium_1999}, and transport sources of energy into the cell \citep{mueckler_slc2_2013}. As with all physical processes, transporters must comply with the principles of thermodynamics \citep{oster_network_1973,hwang_nonequilibrium_2004,beard_chemical_2008}. Because each chemical species within a solution is associated with a chemical potential that increases with concentration, passive transporters can only move substrates from a region of high concentration to a region of low concentration \citep{keener_mathematical_2009,beard_chemical_2008}. To move a substrate against a concentration gradient, a source of energy must be provided. For example, active transporters use ATP hydrolysis to drive the transport of substrates against a potential gradient \citep{keener_mathematical_2009}.

Mathematical models of membrane transporters have been developed for the purpose of understanding their transport mechanism, and predicting their behaviour beyond experimental measurements. However, despite the wealth of models available for transporters, thermodynamic consistency has not usually been applied. As a result, many of these models describe physically infeasible systems where for example, species are transported against their potential gradients in the absence of an energy source, therefore generating energy out of nowhere. In the current literature, thermodynamic consistency is commonly violated through the use of equations that  describe irreversible transporters, neglect dependence on certain metabolites, or have incorrect equilibrium points \citep{gawthrop_energy-based_2014,smith_development_2004,tran_thermodynamic_2009}. While there are methods for incorporating thermodynamic constraints such as detailed balance \citep{liebermeister_modular_2010,beard_chemical_2008,smith_development_2004} and the Nernst potential \citep{keener_mathematical_2009}, they tend to be scattered around the literature, and are not universally applied. Furthermore, many transporters are electrogenic, therefore there is an interaction between chemical and electrical potential \citep{smith_development_2004,terkildsen_balance_2007}, and the multidomain nature of these transporters can confound efforts to develop thermodynamically consistent models. In some cases, thermodynamic inconsistency may impact on the ability of a model to remain physiological under a wide range of conditions. For example, in the context of heart failure, where ATP is depleted, active transporters such as the Na\textsuperscript{+}/K\textsuperscript{+} ATPase and sarcoplasmic/endoplasmic Ca\textsuperscript{2+} ATPase (SERCA) operate at reduced rates. Whole cell models of cardiac cells generally do not describe this metabolite dependence because they neglect thermodynamic constraints \citep{smith_development_2004}, and while thermodynamically consistent models capture these effects \citep{terkildsen_balance_2007,tran_thermodynamic_2009}, they tend to be the exception rather than the rule.

To facilitate the incorporation of thermodynamic constraints into models of membrane transporters, we require a framework that is (a) thermodynamically consistent by design; (b) able to reproduce known thermodynamic constraints; and (c) general enough to model a wide range of transporters. We propose the use of bond graphs, which are a physics-based framework in which energy transfer is explicitly modelled through connections between physical components, therefore satisfying thermodynamic consistency \citep{gawthrop_metamodelling:_1996,borutzky_bond_2010}. The bond graph representation is also domain-independent, therefore it is general enough to represent a wide range of physical systems. Bond graphs were originally invented by Henry Paynter for use in hydroelectric systems \citep{paynter_analysis_1961}, but they also naturally represent electrical and mechanical systems \citep{borutzky_bond_2010}. The reader is referred to the texts by \citet{gawthrop_metamodelling:_1996}, \citet{borutzky_bond_2010} and \citet{gawthrop_bond-graph_2007} for a comprehensive introduction to bond graph theory. More recently, bond graphs have been extended to chemical \citep{thoma_modelling_2000}, biochemical \citep{oster_network_1973,gawthrop_energy-based_2014} and electrochemical systems \citep{gawthrop_bond_2017}, enabling bond graph modelling of membrane transporters such as the sodium-glucose transport protein 1 (SGLT1) \citep{gawthrop_energy-based_2017}. Models of electrogenic membrane transporters have recently been incorporated into a bond graph model of cardiac electrophysiology \citep{pan_bond_2018-1}. A further advantage of modelling transporters as bond graphs is that they are modular, therefore individual models of transporters are easily coupled to other bond graph models to build comprehensive models of biological systems \citep{gawthrop_hierarchical_2015}.

In this paper, we review the bond graph theory required for modelling ion transporters and use bond graphs to build simple hypothetical models of transporters (\autoref{sec:examples}), demonstrating that bond graphs capture important thermodynamic concepts such as detailed balance, free energy of reaction and the Nernst potential. We then develop a bond graph model of cardiac SERCA based on existing work by \citet{tran_thermodynamic_2009}, and used the bond graph model to assess energy consumption and efficiency (\autoref{sec:SERCA}). Bond graphs also provide a framework for detecting thermodynamic inconsistencies within existing models. In \autoref{sec:NaK_pump} we develop a new model of Na\tsp{}/K\tsp{} ATPase based on existing work by \citet{terkildsen_balance_2007}, and verify that the model complies with thermodynamic constraints. These examples illustrate that bond graphs are a unifying framework for accounting for thermodynamic constraints in models of membrane transporters. We believe that the bond graph approach will prove to be a powerful tool in the development of thermodynamically consistent models of transporters, and other cellular processes in which energy transduction plays an important role.

\section{Hypothetical models}
\label{sec:examples}
\subsection{Enzyme cycle}
\label{sec:enzyme_cycle}
\begin{figure}
	\centering
	\includegraphics[width=\linewidth]{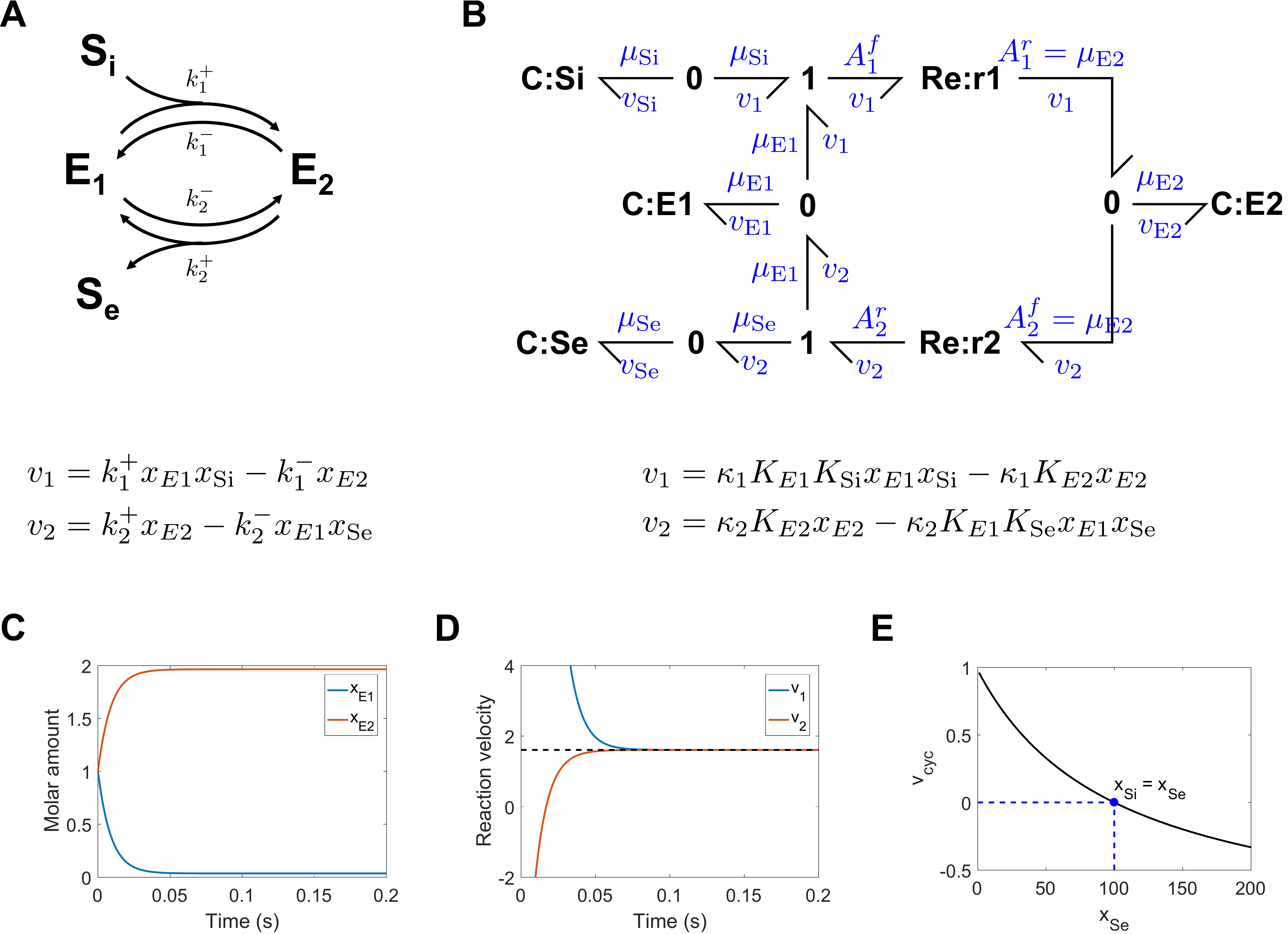}
	\caption{\textbf{A simple enzyme cycle.} \textbf{(A)} Kinetic model; \textbf{(B)} Bond graph model. $\mathrm{S_i}$ and $\mathrm{S_e}$ are the intracellular and extracellular substrate respectively, $E_1$ is the unbound transporter, and $E_2$ is the transporter bound to substrate. A simulation was run with unity parameters, the initial conditions $x_{E1} = x_{E2} = 1$ and $x_\mathrm{Si} = 100$ and $x_\mathrm{Se} = 10$. The resulting molar amounts of transporter states \textbf{(C)} and reaction velocities \textbf{(D)} were plotted against time. \textbf{(E)} Using $x_\mathrm{Si} = 100$, the steady-state cycling rates were plotted against $x_\mathrm{Se}$, with the blue dot indicating the equilibrium point.}
	\label{fig:enzyme_cycle}
\end{figure}

In biochemical systems, Gibbs free energy (also called Gibbs energy or free energy) is transmitted between species and reactions \citep{atkins_physical_2006}. Like all physical systems, biochemical systems must comply with the laws of thermodynamics, therefore reactions can only proceed in the direction of decreasing chemical potential. However, chemical potential is ignored by many existing models, often resulting in physically unrealistic behaviour \citep{gawthrop_energy-based_2014,gawthrop_hierarchical_2015}. Bond graphs account for chemical potential by decomposing the transfer of power into a product of chemical potential ($\mu$ [$\si{J/mol}$]) and molar flow rate ($v$ [$\si {mol/s}$]) \citep{gawthrop_energy-based_2014}. Power is transmitted between components within a bond graph by connecting them with bonds that carry chemical potential and molar flow rate. Because the power leaving one component is transmitted to another, bond graphs are thermodynamically consistent, that is, the flow of energy is explicitly accounted for. In bond graph terminology, chemical potential is known as the effort variable, and molar flow rate is known as the flow variable \citep{gawthrop_energy-based_2014}.

In this section, we use a simple transporter model to outline the essential aspects of modelling biochemical systems using bond graphs (\autoref{fig:enzyme_cycle}A). This transporter model contains a two-state enzyme cycle, where an uncharged substrate S binds to the enzyme on the intracellular (i) side, and unbinds on the extracellular (e) side. We denote the substrate in a specific compartment as $\mathrm{S}_x$, where $x \in \{ \mathrm{i}, \mathrm{e} \}$. The kinetic model is shown in \autoref{fig:enzyme_cycle}A, and the bond graph representation of this reaction cycle is shown in \autoref{fig:enzyme_cycle}B. Because a thermodynamic approach implies reversibility for all reactions \citep{polettini_irreversible_2014,gawthrop_energy-based_2014}, the transport of the substrate can proceed in both directions, depending on the direction of the chemical gradient. 

Chemical energy is stored within the species of a biochemical system \citep{atkins_physical_2006}. This energy allows the system to do work through chemical reactions. The ability of a species $s$ to drive a reaction is given by the chemical potential $\mu_s$ [J/mol], which is dependent on the molar amount of species $x_s$ [mol] in a logarithmic manner \citep{keener_mathematical_2009,atkins_physical_2006}:
\begin{align}
\mu_s = RT\ln(K_s x_s)
\label{eq:cp}
\end{align}
where $R=8.314\ \si{J/K/mol}$ is the ideal gas constant, $T$ [K] is the absolute temperature and $K_s$ [mol$^{-1}$] is the species thermodynamic constant. Therefore, the chemical potentials of the species in the enzyme cycle are
\begin{align}
\mu_\text{Si} &= RT\ln(K_\text{Si}x_\text{Si}) \\
\mu_\text{Se} &= RT\ln(K_\text{Se}x_\text{Se}) \\
\mu_\text{E1} &= RT\ln(K_\text{E1}x_\text{E1}) \\
\mu_\text{E2} &= RT\ln(K_\text{E2}x_\text{E2})
\end{align}
Because species store energy, they are represented as \textbf{C} components in the bond graph model (\autoref{fig:enzyme_cycle}B), as they are analogous to capacitors in electrical circuits.

The chemical energy stored within the various species of a biochemical system is used to drive reactions, which convert chemical species into other chemical species, dissipating chemical energy in the process. The rate of reaction can be related to the chemical potential of its reactants and products using the Marcelin-de Donder equation \citep{gawthrop_bond-graph_2007,oster_network_1973}:
\begin{align}
v_R = \kappa_R (e^{A_R^f /RT} - e^{A_R^r /RT})
\end{align}
where for a reaction $R$, $v_R$ [mol/s] is the rate of reaction, $\kappa_R$ [mol/s] is the reaction rate constant, $A_R^f$ [J/mol] is the forward affinity (representing the total chemical potential of the reactants) and $A_R^r$ [J/mol] is the reverse affinity (representing the total chemical potential of the products). For the enzyme cycle example in \autoref{fig:enzyme_cycle}B, the reaction rates of the upper and lower reactions are given by the equations
\begin{align}
v_1 = \kappa_1 (e^{A_1^f /RT} - e^{A_1^r /RT}) \label{eq:enzyme_r1} \\
v_2 = \kappa_2 (e^{A_2^f /RT} - e^{A_2^r /RT}) \label{eq:enzyme_r2}
\end{align}
As seen in \autoref{fig:enzyme_cycle}B, reactions are represented as \textbf{Re} components in bond graph modelling.

In the enzyme cycle, some species are involved in more than one reaction. Therefore we require constraints to ensure that (a) the same chemical potential is used for the involvement of the same species in different reactions and (b) the contributions of each reaction are summed when calculating the rate of change of each species. These constraints are captured by the \textbf{0} junction in bond graph modelling, which accounts for the first constraint by setting the chemical potentials of all connected bonds to be equal. In \autoref{fig:enzyme_cycle}B, there are four \textbf{0} junctions corresponding to each species. The second constraint is captured through the conservation equations
\begin{align}
	v_\text{Si} &= -v_1 \label{eq:enzyme_0_start} \\
	v_\text{Se} &= v_2 \\
	v_\text{E1} &= v_2 - v_1 \\
	v_\text{E2} &= v_1 - v_2 \label{eq:enzyme_0_end} 
\end{align}
\textbf{0} junctions are analogous to parallel connections in electric circuits, and the Eqs. \ref{eq:enzyme_0_start}--\ref{eq:enzyme_0_end} represent the biochemical version of Kirchhoff's current law. Note that the molar flow rates of the bonds connecting to each \textbf{0} junction sum to zero, therefore the \textbf{0} junction conserves power \citep{borutzky_bond_2010}.

In this example, both reactions involve the combination or dissociation of chemical species, and therefore have multiple reactants or products. To account for this, we require constraints so that (a) the rate at which each species is produced/consumed by the reaction is equal to the rate of reaction and (b) the affinities of each reaction are the totals of the chemical potentials of the reactants or products. These constraints are captured by \textbf{1} junctions in bond graph modelling, which satisfy the first constraint by fixing the molar flow rates of each connected bond to equal values. In \autoref{fig:enzyme_cycle}B, there are two \textbf{1} junctions representing the association and dissociation of substrate. The second constraint is accounted for through the conservation laws
\begin{align}
	A_1^f = \mu_\text{Si} + \mu_\text{E1} \label{eq:enzyme_A1f} \\
	A_2^r = \mu_\text{Se} + \mu_\text{E1} \label{eq:enzyme_A2r}
\end{align}
\textbf{1} junctions are analogous to series connections in electric circuits, and Eqs. \ref{eq:enzyme_A1f}--\ref{eq:enzyme_A2r} are biochemical versions of Kirchhoff's voltage law. Note that since the chemical potentials of all bonds connected to \textbf{1} junctions sum to zero, they are power conserving, like the \textbf{0} junction \citep{borutzky_bond_2010}.

Using the components described above, it is possible to derive a differential equation for the enzyme cycle. By substituting Eqs. \ref{eq:enzyme_A1f}--\ref{eq:enzyme_A2r} into Eqs. 
\ref{eq:enzyme_r1}--\ref{eq:enzyme_r2}, we find that the reaction rates are
\begin{align}
v_1 &= \kappa_1 (e^{(\mu_\text{Si} + \mu_\text{E1})/RT} - e^{\mu_\text{E2} /RT}) 
= \kappa_1 K_\text{Si} K_\text{E1} x_\text{Si} x_\text{E1} - \kappa_1 K_\text{E2} x_\text{E2}  \\
v_2 &= \kappa_2 (e^{\mu_\text{E2} /RT} - e^{(\mu_\text{Se} + \mu_\text{E1}) /RT}) 
= \kappa_2 K_\text{E2} x_\text{E2} - \kappa_2 K_\text{Se} K_\text{E1} x_\text{Se} x_\text{E1}
\end{align}
Hence the bond graph components are able to reproduce the mass action equations, with the kinetic parameters
\begin{align}
k_1^+ &= \kappa_1 K_{E1} K_\mathrm{Si} \label{eq:kinetic_bg_enzyme_start} \\
k_1^- &= \kappa_1 K_{E2} \\
k_2^+ &= \kappa_2 K_{E2} \\
k_2^- &= \kappa_2 K_{E1} K_\mathrm{Se}
\label{eq:kinetic_bg_enzyme}
\end{align}
By using the conservation laws from Eqs. \ref{eq:enzyme_0_start}--\ref{eq:enzyme_0_end}, the rates of change for each species are
\begin{align}
\dot{x}_\text{Si}
&= v_\text{Si} = -v_1 \nonumber \\
&= -k_1^+ x_\text{Si} x_\text{E1} + k_1^- x_\text{E2} \label{eq:enzyme_Si} \\
\dot{x}_\text{Se} &= v_\text{Se} = v_2   \nonumber \\
&= k_2^+ x_\text{E2} - k_2^- x_\text{Se} x_\text{E1} \label{eq:enzyme_Se} \\
\dot{x}_\text{E1}
&= v_\text{E1} = v_2 - v_1  \nonumber \\
&= -k_1^+ x_\text{Si} x_\text{E1} + k_1^- x_\text{E2}  + k_2^+ x_\text{E2} - k_2^- x_\text{Se} x_\text{E1} \\
\dot{x}_\text{E2}
&= v_\text{E2} = v_1 - v_2  \nonumber\\
&= k_1^+ x_\text{Si} x_\text{E1} - k_1^- x_\text{E2}  - k_2^+ x_\text{E2} + k_2^- x_\text{Se} x_\text{E1}
\end{align}

The above equations are applicable to an isolated system. However, since transporters are dissipative systems, i.e. they release energy as heat, it is impossible for these to operate continuously at steady state without an external power supply. Therefore, we model these external power supplies by holding certain species at constant concentrations rather than allowing them to dynamically vary. Such species are known as chemostats \citep{polettini_irreversible_2014}. In this example, we assume that the concentrations of the substrates $\mathrm{S_i}$ and $\mathrm{S_e}$ are constant, so that Eqs. \ref{eq:enzyme_Si} and \ref{eq:enzyme_Se} are replaced by
\begin{align}
	\dot{x}_\text{Si} = 0 \\
	\dot{x}_\text{Se} = 0
\end{align}
Because an external flow of species is required to keep the concentration of the species constant, chemostats are not energy conserving, but rather represent the influx or release of energy into the external environment. Thus chemostats turn a biochemical system from an isolated system into an open system that communicates with its external environment, and can therefore be used to couple models together. In bond graph modelling, \textbf{C} components are replaced with \textbf{Se} (effort source) components when they become chemostats to indicate the transfer of power with an external source. Since species can be represented using \textbf{C} or \textbf{Se} components interchangeably depending on the purpose of the model \citep{gawthrop_bond_2017-1,gawthrop_bond_2017}, we represent species as \textbf{C} components in all diagrams within this paper, and mention in text whether certain species are treated as ``chemostats'' in the analysis.

For the remainder of \autoref{sec:examples}, we assume that:
\begin{enumerate}
	\item Apart from the states of the transporter (e.g. $\mathrm{E_1}$, $\mathrm{E_2}$), the concentrations of all species are constant (i.e. they are modelled as chemostats).
	\item The volumes of all compartments are equal, so that the amount of each species directly corresponds to concentration. In real biological systems, compartments will have different volumes, and we show how to incorporate these effects in Appendix B of the Supplementary Material.
	\item The parameters $K$ and $\kappa$ for each species and reaction take on unity values.
\end{enumerate}
Assumptions 2 and 3 are made to simplify the analysis in this section, although we show how to generalise beyond these assumptions in later sections.

\autoref{fig:enzyme_cycle}C,D shows simulations of this model when the amount of enzyme is small relative to the amount of substrate, thus we expect the system to achieve a steady state relatively quickly. At steady state, the amounts of each pump state are constant, and the two reaction velocities converge towards the same value, as dictated by the pathway analysis of \citet{gawthrop_energy-based_2017}. We are often interested in the cycling (or turnover) rate at steady state, $v_\text{cyc}$, which is given by $v_\text{cyc} = V/e_0$, where $V$ is the steady-state reaction rate, and $e_0 = x_\text{E1} + x_\text{E2}$ is the total amount of transporter \citep{atkins_physical_2006}.

\autoref{fig:enzyme_cycle}E shows the effect of changing $x_\mathrm{Se}$ on cycling rate, indicating an inverse relationship: as $x_\mathrm{Se}$ increases, the transporter operates at a lower rate, and eventually operates in the reverse direction as expected. The bond graph model captures the fundamental physical constraint that the equilibrium point between the forward and reverse regimes of operation occurs when $x_\mathrm{Se} = x_\mathrm{Si}$. Therefore the simple transporter can only allow transport of its substrate down its concentration gradient.

The direction in which each reaction proceeds is determined by the Gibbs free energy of reaction $\Delta G$. Because reactions can only run in the direction of decreasing chemical potential, they will only proceed in the forward direction if $\Delta G$ is negative. The Gibbs free energy of reaction relates to the affinities $A = A^f - A^r$ of the \textbf{Re} components in bond graph models. Since efforts in the biochemical domain are associated with Gibbs free energy, the Gibbs free energy of reaction is the negative of affinity:
\begin{align}
\Delta G = -A = A^r - A^f
\end{align}
Therefore, the free energies of each of the reactions in this example are
\begin{align}
\Delta G_1 &= \mu_\text{E2} - \mu_\text{E1} - \mu_\text{Si} \\
\Delta G_2 &= \mu_\text{E1} + \mu_\text{Se} - \mu_\text{E2}
\end{align}

At steady state, the transporter's direction of operation is determined by the Gibbs free energy of the overall reaction $S_\mathrm{i} \rightleftharpoons \mathrm{S_e}$, which is also the sum of all reactions in its cycle \citep{gawthrop_energy-based_2017}:
\begin{align}
\Delta G = \Delta G_1 + \Delta G_2  = \mu_\mathrm{Se} - \mu_\mathrm{Si}
\end{align}
By substituting Eq. \ref{eq:cp} and setting $\Delta G = 0$, the equilibrium of the system can be found:
\begin{align}
\Delta G &= RT\ln (K_\mathrm{Se} x_\mathrm{Se})  - RT\ln (K_\mathrm{Si} x_\mathrm{Si}) 
= RT \ln \left(  x_\mathrm{Se} / x_\mathrm{Si}  \right) = 0 \\
& \Rightarrow  x_\mathrm{Se}  = x_\mathrm{Si}
\end{align}
Therefore, as expected, the point at which the free energy is zero corresponds to the equilibrium of the transporter.

To ensure that the equilibrium occurs at $x_\mathrm{Se} = x_\mathrm{Si}$, we must specify the equilibrium between the substrate in each compartment:
\begin{align}
K_\mathrm{Si}/K_\mathrm{Se} = K^{\text{eq}} = 1
\label{eq:detailed_balance_simple}
\end{align}
which is exactly equivalent to the detailed balance constraint used in kinetic models \citep{keener_mathematical_2009,smith_development_2004,tran_thermodynamic_2009}:
\begin{align}
\frac{k_1^+ k_2^+}{k_1^- k_2^-} = 1
\label{eq:detailed_balance_kinetic_enzyme}
\end{align}
(this can be easily checked by substituting Eq. \ref{eq:detailed_balance_simple} into Eqs. \ref{eq:kinetic_bg_enzyme_start}--\ref{eq:kinetic_bg_enzyme}). Because bond graphs are thermodynamically consistent, the thermodynamic constraint is simpler and more intuitive when compared to that for the kinetic parameters.

\subsection{Coupled reactions}
\begin{figure}
	\centering
	\includegraphics[width=\linewidth]{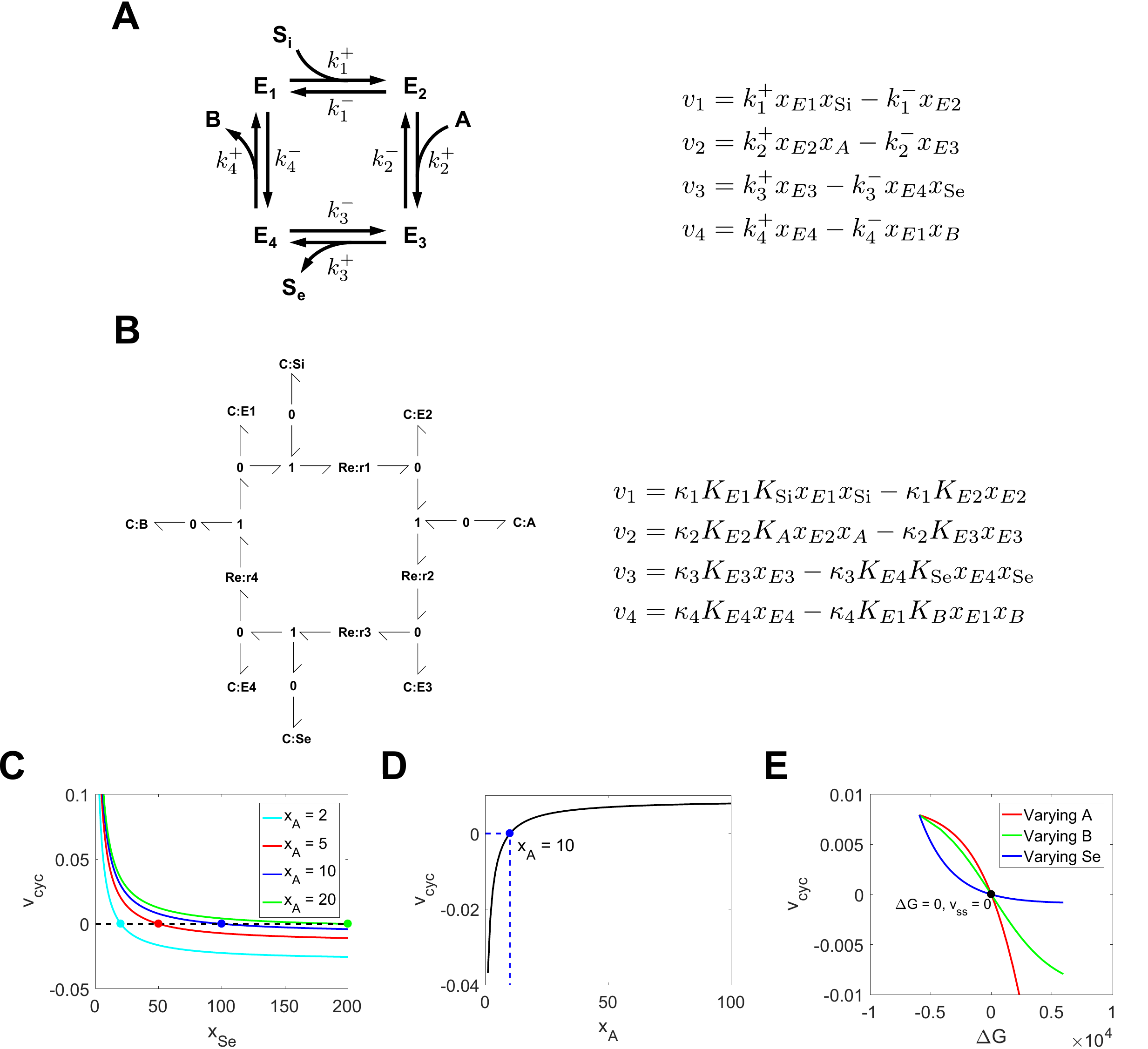}
	\caption{\textbf{Transport of a species by coupling to an energy supply.} \textbf{(A)} Kinetic model; \textbf{(B)} Bond graph model. $\mathrm{S_i}$ and $\mathrm{S_e}$ are the intracellular and extracellular substrate respectively, $A$ and $B$ are species that provide power for the transport cycle, and $E_1$, $E_2$, $E_3$ and $E_4$ are the transporter states. Simulations were run using the initial conditions $x_{E1} = x_{E2} = x_{E3} = x_{E4} = 0.5$, and chemostats $x_B = 1$, $x_\mathrm{Si} = 10$. \textbf{(C)} The effect of $x_\mathrm{Se}$ on steady state cycling rate for different $x_A$. Equilibrium points, where the cycling rates are zero, are denoted by the dots. \textbf{(D)} Effect of $x_A$ on the cycling rate, with $x_\mathrm{Se} = 100$. The equilibrium (blue dots) occurs at $x_A = x_\mathrm{Se}/(x_\mathrm{Si}x_B) = 10$. \textbf{(E)} The relationship between the Gibbs free energy of the transporter and cycling rate, when A, B and $\mathrm{S_e}$ are varied.}
	\label{fig:coupled_reactions}
\end{figure}
In the previous example, we observed that a passive transporter was unable to move a substrate against its concentration gradient. However, many transporters, including active transporters and exchangers, are able to move substrates against their concentration gradients. Such transporters couple the movement of substrate against a concentration gradient to a process that provides sufficient energy to enable transport. In \autoref{fig:coupled_reactions}A, we show a simple mechanism for coupling together these processes. In addition to binding $\mathrm{S_i}$ and unbinding $\mathrm{S_e}$, the transport also binds $A$ and unbinds $B$, giving rise to the overall reaction
\begin{align}
	\mathrm{S_i} + A \rightleftharpoons \mathrm{S_e} + B
\end{align}
The reaction $A \rightleftharpoons B$ provides energy for the transport of substrate, and may for example represent ATP hydrolysis in active transporters, or the transport of another species down its concentration gradient in exchangers or cotransporters. The bond graph representation of this chemical reaction network is given in \autoref{fig:coupled_reactions}B.

To achieve the correct equilibrium point, we use the following thermodynamic constraint, describing the equilibria between side species:
\begin{align}
	K_\mathrm{Si}/K_\mathrm{Se} &= 1 \\
	K_A/K_B &= 1
\end{align}
Note that $K_A/K_B$ can be set to different values depending on the process that $A \rlh B$ represents; it has a value of 1 if it represents the transport of an uncharged substrate (as the equilibrium point occurs when the substrate has the same concentration on either side of the membrane), whereas it is determined by the standard Gibbs free energy of reaction if it represents a reaction such as ATP hydrolysis (see Appendix B of the Supplementary Material).

For a kinetic model with the parameters $k_1^+$, $k_1^-$, $k_2^+$, $k_2^-$, $k_3^+$, $k_3^-$, $k_4^+$ and $k_4^-$, the corresponding thermodynamic constraint is
\begin{align}
	\frac{k_1^+ k_2^+ k_3^+ k_4^+}{k_1^- k_2^- k_3^- k_4^-} = \frac{K_\mathrm{Si}}{K_\mathrm{Se}} \frac{K_A}{K_B} = 1
\end{align}
We note here that the thermodynamic constraint between kinetic parameters is more complicated than that in \autoref{sec:enzyme_cycle} because there are more reactions in the cycle. When the number of reactions in biochemical cycles increases, these thermodynamic constraints become more complex, and harder to derive. By contrast, when bond graph parameters are used, thermodynamic constraints remain simple regardless of cycle complexity, and are more intuitive to formulate.

In \autoref{fig:coupled_reactions}C we simulated the bond graph model to show the steady-state cycling rates under different values of $x_A$ and $x_\mathrm{Se}$. We note that for a passive transporter the equilibrium point occurs at $x_\mathrm{Se} = x_\mathrm{Si} = 10$, however in this example with $x_A > x_B$, the equilibrium points (represented by dots) occur at $x_\mathrm{Se} > x_\mathrm{Si}$. With all thermodynamic parameters set to 1, the equilibrium is given by the equation
\begin{align}
	x_\mathrm{Se} = x_\mathrm{Si} x_A/x_B
	\label{eq:coupled_eq}
\end{align}
Thus when $x_A > x_B$ (i.e. the coupled process releases energy), $x_\mathrm{Se}$ is greater than $x_\mathrm{Si}$ at equilibrium. Therefore there is a region of operation (when $x_\mathrm{Si} < x_\mathrm{Se} < x_\mathrm{Si} x_A/x_B$) where the transporter can transport S against its concentration gradient. As the value of $x_A$ increases, the equilibrium point shifts towards the right (\autoref{fig:coupled_reactions}C), indicating that the transporter is able to move S against a greater concentration gradient. This is because with a greater $x_A$, the reaction $A \rightleftharpoons B$ provides more power, driving the transport in the forward direction (\autoref{fig:coupled_reactions}D).

The Gibbs free energies of each of the reactions are
\begin{align}
	\Delta G_1 &= \mu_\text{E2} - \mu_\text{E1} - \mu_\text{Si} \\
	\Delta G_2 &= \mu_\text{E3} - \mu_\text{E2} - \mu_A \\
	\Delta G_3 &= \mu_\text{E4} + \mu_\text{Se} - \mu_\text{E3}  \\
	\Delta G_4 &= \mu_\text{E1} + \mu_B - \mu_\text{E4}
\end{align}
and the free energy of the transporter is
\begin{align}
	\Delta G = \Delta G_1 + \Delta G_2 + \Delta G_3 + \Delta G_4 = \mu_\mathrm{Se} + \mu_B - \mu_\mathrm{Si} - \mu_A
\end{align}
By substituting Eq. \ref{eq:cp} and setting $\Delta G = 0$, we can recover the equilibrium relationship in Eq. \ref{eq:coupled_eq}:
\begin{align}
\Delta G &= RT\ln (K_\mathrm{Se} x_\mathrm{Se}) + RT\ln (K_B x_B) - RT\ln (K_\mathrm{Si} x_\mathrm{Si}) - RT\ln (K_A x_A) \nonumber \\
&= RT \ln \left( \frac{K_\mathrm{Se} x_\mathrm{Se} K_B x_B}{K_\mathrm{Si} x_\mathrm{Si} K_A x_A} \right)\nonumber  \\
&= RT \ln \left( \frac{ x_\mathrm{Se}  x_B}{ x_\mathrm{Si} x_A} \right) = 0 \\
& \Rightarrow \frac{ x_\mathrm{Se}  x_B}{ x_\mathrm{Si}x_A}  = 1
\end{align}
Therefore, the equilibrium for steady-state operation corresponds to the point where the Gibbs free energy of the transport process is zero. We verify this fundamental physical constraint by plotting Gibbs free energy against cycling rate (\autoref{fig:coupled_reactions}E). We varied the concentrations of A, B and $\mathrm{S_e}$ to generate three different curves. Despite differences in the shape of each curve, they each pass through the equilibrium point $\Delta G = 0$, $v_\text{cyc} = 0$, verifying that the bond graph model correctly captures the relationship between Gibbs free energy and equilibrium. Furthermore, the transporter operates in the forward direction ($v_\text{cyc} > 0$) with a negative (favourable) Gibbs free energy, and in the reverse direction ($v_\text{cyc} < 0$) with a positive (unfavourable) free energy.

\subsection{Electrogenic transport}
\label{sec:charged_example}
\begin{figure}
	\centering
	\includegraphics[width=\linewidth]{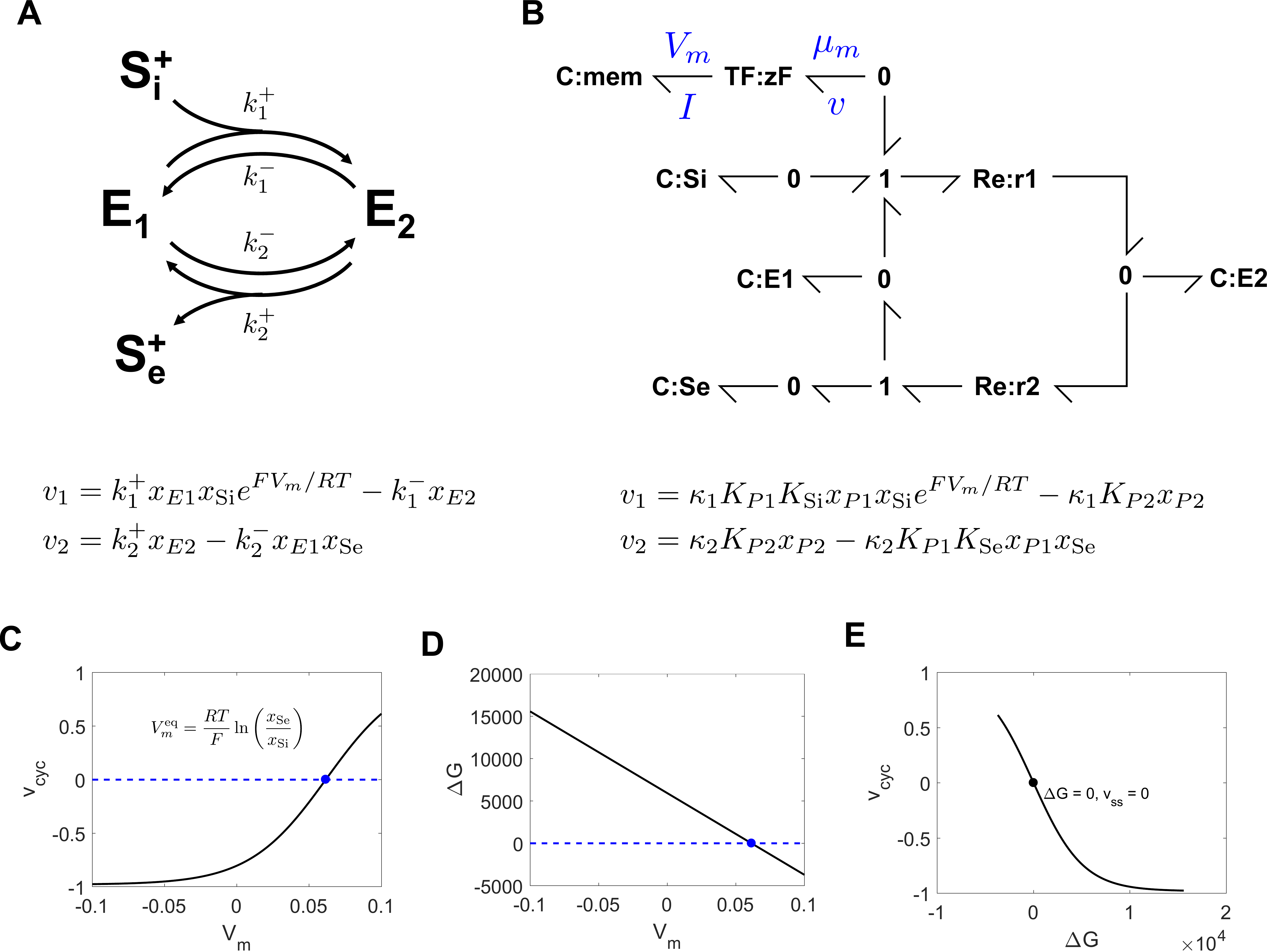}
	\caption{\textbf{Transport of a charged species} \textbf{(A)} Kinetic model; \textbf{(B)} Bond graph model. $\mathrm{S_i^+}$ and $\mathrm{S_e^+}$ are the intracellular and extracellular substrate respectively, $E_1$ is the unbound transporter, and $E_2$ is the transporter bound to substrate. Simulations were run with the initial conditions $x_{E1} = x_{E2} = 1$, and chemostats $x_\mathrm{Si} = 10$, $x_\mathrm{Se} = 100$. \textbf{(C)} Relationship between the membrane potential and cycling rate. The blue dot indicates the equilibrium potential, as predicted by the Nernst equation. \textbf{(D)} The effect of membrane potential on the Gibbs free energy of the transporter. \textbf{(E)} Relationship between the Gibbs free energy and cycling rate under the simulation conditions, with varying membrane potential.}
	\label{fig:charged_species}
\end{figure}

Many membrane transporters, including ion exchangers, cotransporters and active ion transporters, move a charged species across a membrane. For those membranes within the cell (such as the plasma membrane, and inner mitochondrial membrane) that are charged, the membrane potential influences both the kinetics and thermodynamics of transport. In this section we explore the impact on transport by modifying the enzyme example in \autoref{sec:enzyme_cycle}, so that the transported substrate  has a single unit of positive charge (i.e. $z=1$) (\autoref{fig:charged_species}A). For simplicity, we choose to assign the entirety of the electrical dependence to the forward side of the first reaction, which results in an exponential dependence on membrane voltage, arising from thermodynamic constraints \citep{keener_mathematical_2009,smith_development_2004}.

Because the bond graph representation is domain-independent, physical processes from other physical domains such as electrical or mechanical domains can be modelled by changing the units of the effort and flow variables. In the bond graph model (\autoref{fig:charged_species}B), the electrical dependence is incorporated by adding a \textbf{C} component representing membrane capacitance. This \textbf{C} component has similar energy-storage properties to the \textbf{C} component for each chemical species, however, in place of Eq. \ref{eq:cp}, it has a linear constitutive equation $V_m = q_m/C$, where $V_m$ [V] is the membrane potential and $q_m$ [C] is the charge. We use a membrane capacitance of $C=F^2$. Since the capacitor is an electrical component, its power must be converted to chemical power to appropriately describe its influence on reaction kinetics. This  conversion is given by Faraday's constant $F=96485\ \si{C/mol}$, which relates the membrane potential and current to chemical potential and molar flow rate \citep{gawthrop_bond_2017-1}:
\begin{align}
	\mu_m &= zFV_m \\
	I &= zFv
\end{align}
These transformations are described by the \textbf{TF} component in \autoref{fig:charged_species}B. Since $\mu_m v = V_m I$, the \textbf{TF} component is a power-conserving transformation.

By substituting the relevant constitutive equations, the rate of reaction 1 is given by
\begin{align}
v_1 = \kappa_1 (e^{(\mu_\mathrm{Si} + \mu_{E1} + \mu_m)/RT} - e^{\mu_{E2}/RT}) = \kappa_1 K_\mathrm{Si} K_{E1} x_\mathrm{Si} x_{E1} e^{FV_m/RT} - \kappa_1 K_{E2} x_{E2}
\end{align}
which corresponds to the exponential dependence of the kinetic reaction scheme. Because adding an electrical component affects the thermodynamics of transport, the equilibrium becomes dependent on membrane potential. As a result, the kinetics of the transporter must depend on membrane voltage to account for changes in equilibrium, although in practice modellers often make assumptions about where the electrical dependence lies due to a lack of experimental data \citep{smith_development_2004}.

Bond graphs incorporate thermodynamic constraints for electrogenic transport. Because the chemical reaction structure is similar to the system in \autoref{sec:enzyme_cycle}, the constraint in Eq. \ref{eq:detailed_balance_simple} holds. However, because a single unit of charge is moved across the membrane in a single cycle, there is an additional constraint for charge:
\begin{align}
	z_1^f - z_1^r + z_2^f - z_2^r = z = 1
	\label{eq:electrogenic_charge}
\end{align}
where $z$ indicates charge, subscripts indicate the reaction number, and superscripts indicate the side of the reaction. In this example, $z_1^f = 1$ and $z_1^r = z_2^f = z_2^r = 0$, therefore Eq. \ref{eq:electrogenic_charge} is satisfied.

At equilibrium, the free energy of the overall reaction is zero \citep{gawthrop_energy-based_2017}, and because $K_\mathrm{Si} = K_\mathrm{Se}$ (Eq. \ref{eq:detailed_balance_simple}),
\begin{align}
\Delta G = \mu_\mathrm{Se} - \mu_\mathrm{Si} - \mu_m = RT\ln (K_\mathrm{Se} x_\mathrm{Se}) - RT\ln (K_\mathrm{Si} x_\mathrm{Si}) - FV_m = 0
\label{eq:DeltaG_electrogenic}
\end{align}
the membrane potential at equilibrium is
\begin{align}
V_m = \frac{RT}{F} \ln \left( \frac{x_\mathrm{Se}}{x_\mathrm{Si}} \right)
\label{eq:Nernst}
\end{align}
which is the familiar Nernst potential \citep{keener_mathematical_2009}.

In \autoref{fig:charged_species}C, this system was simulated under varying membrane potentials. Because the species is positively charged, changing the membrane potential in the positive direction drives transport of the species out of the cell. The transporter achieved an equilibrium at a membrane potential of approximately 0.062, consistent with the Nernst equation in Eq. \ref{eq:Nernst}.

In \autoref{fig:charged_species}D we plot the Gibbs free energy (defined in Eq. \ref{eq:DeltaG_electrogenic}) against membrane potential. The membrane potential has a linear contribution to the free energy of the transporter, becoming more negative for positive membrane potentials due to the positively charged substrate. Importantly, the zero of the Gibbs free energy coincides with the zero of the cycling rate in \autoref{fig:charged_species}C, thus the transporter is dissipative, and therefore thermodynamically consistent. \autoref{fig:charged_species}E shows this directly, where the transporter operates in the positive direction only if the Gibbs free energy is negative.

\section{Thermodynamic models of cardiac cell transporters}
\label{sec:results}
\subsection{SERCA}
\label{sec:SERCA}
\begin{figure}
	\centering
	\includegraphics[width=\linewidth]{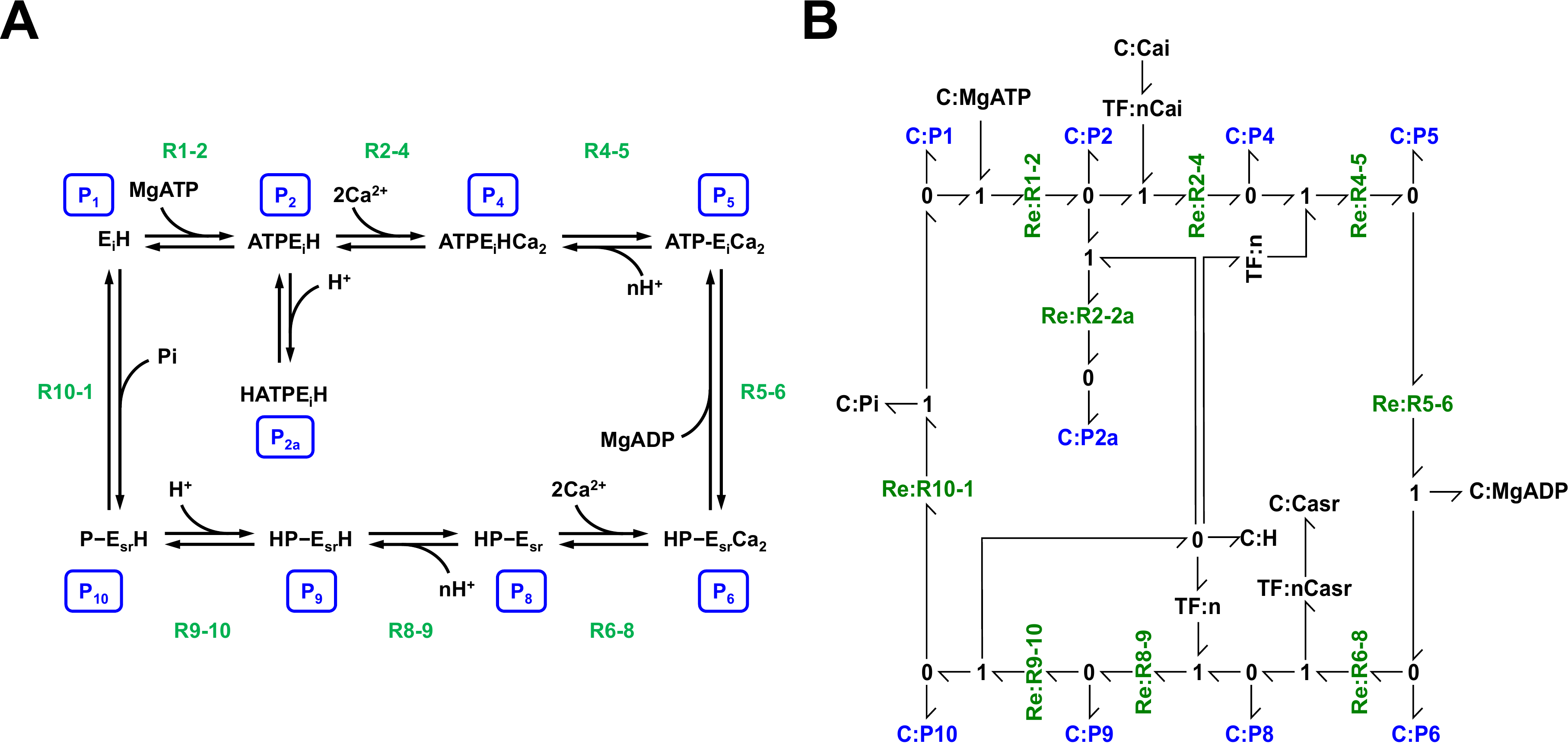}
	\caption{\textbf{Kinetic and bond graph representations of the cardiac SERCA model from \citet{tran_thermodynamic_2009}}. \textbf{(A)} Kinetic model, adapted from Fig. 2 of \citet{tran_thermodynamic_2009}. The numbers for each pump state are shown in blue boxes, and the names for each reaction are labelled in green. \textbf{(B)} Bond graph representation, with pump states shown in blue, and reactions in green.}
	\label{fig:SERCA}
\end{figure}
The sarcoplasmic/endoplasmic Ca\textsuperscript{2+} ATPase (SERCA) is an active ion transporter that pumps calcium from the cytosol into the sarcoplasmic reticulum (SR), an intracellular Ca\textsuperscript{2+} store, restoring the calcium released from the SR with each heart beat. Because SERCA pumps calcium against a concentration gradient ($\mathrm{[Ca^{2+}]_\text{sr}} > \mathrm{[Ca^{2+}]_\text{i}}$), it couples calcium transport to the hydrolysis of ATP in order to obtain the energy required for transport. The overall reaction of the pump is
\begin{align}
2\text{Ca}_\text{i}^{2+} + \text{MgATP} \rightleftharpoons
2\text{Ca}_\text{sr}^{2+} + \text{MgADP} +\text{P}_\text{i} + \text{H}^+ 
\end{align}
Since the SR membrane is uncharged, the pump is driven purely by chemical energy.

SERCA has been included in a wide variety of models of cardiac cell Ca\textsuperscript{2+} cycling, excitation-contraction coupling and electrophysiology. Our analysis here is based on the model by \citet{tran_thermodynamic_2009} which has since been incorporated into a number of subsequent cardiac cell models because it describes the dependence of cycling rate on all metabolites \citep{tran_regulation_2015,williams_dynamics_2011,walker_superresolution_2014}. The chemical reaction network and bond graph representation of the \citet{tran_thermodynamic_2009} model are shown in \autoref{fig:SERCA}. The structure of the bond graph closely resembles the cyclic nature of the chemical reaction scheme. However, the bond graph representation explicitly shows that the H\tsp{} ions involved in multiple reactions are linked. Note that the bond graph representation uses \textbf{TF} components to describe multiple copies of a species in a single reaction; this is discussed further in Appendix A of the Supplementary Material.

\begin{figure}
	\centering
	\includegraphics[width=\linewidth]{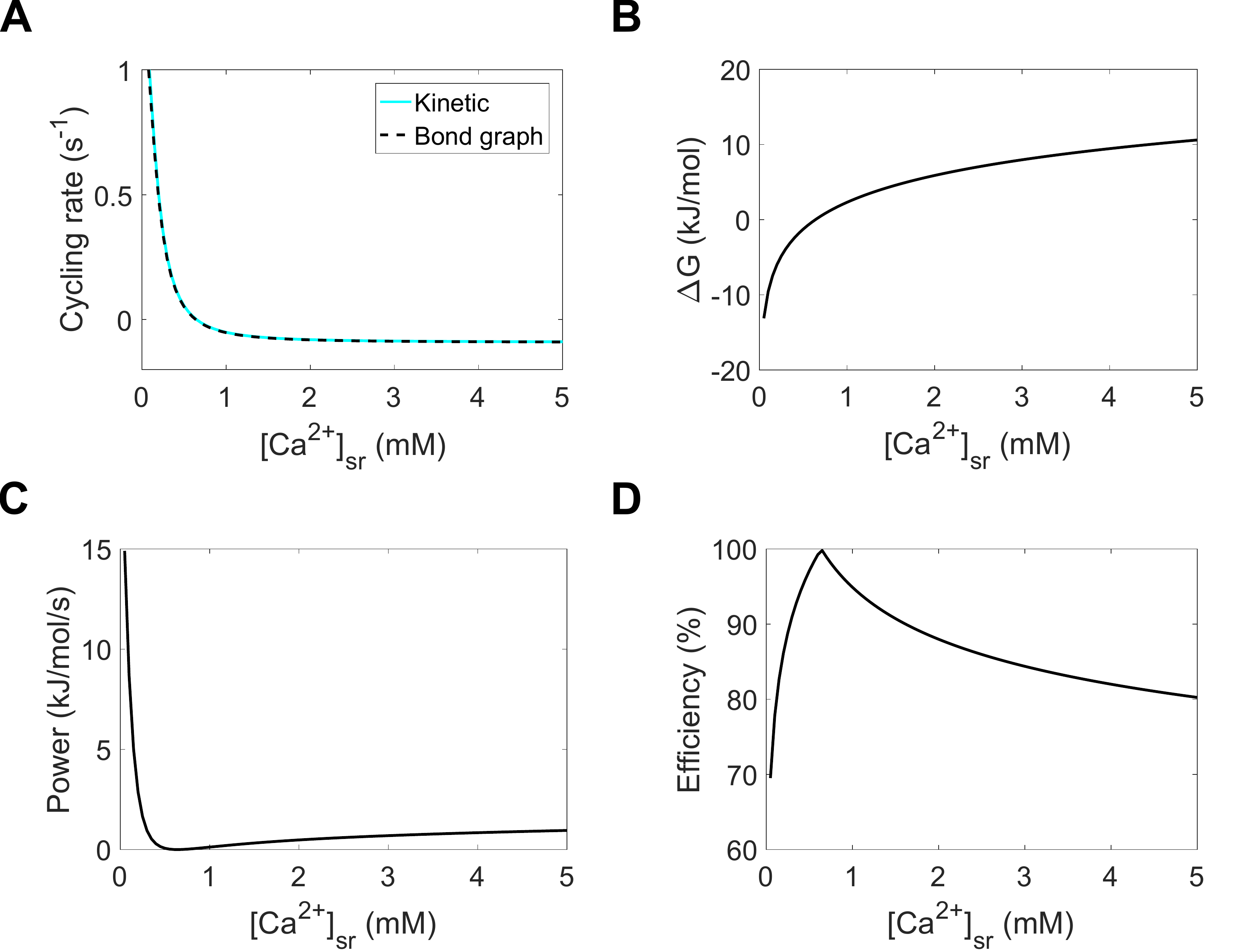}
	\caption{\textbf{Simulation of the SERCA pump.} \textbf{(A)} Comparison of cycling rates for kinetic and bond graph models, reproducing part of Fig. 13 in \citet{tran_thermodynamic_2009}; \textbf{(B)} Gibbs free energy; \textbf{(C)} Power consumption per mol of pump; \textbf{(D)} Pump efficiency. Simulations were run with $\mathrm{[Ca^{2+}]_i = 150 \ \si{nM}}$, $\mathrm{pH = 4}$, $\mathrm{[MgADP] = 0.0363 \ \si{mM}}$, $\mathrm{[MgATP] = 0.1 \ \si{mM}}$, $\mathrm{[Pi] = 15 \ \si{mM}}$. Cycling rates were estimated by initialising each pump state to 1/9 fmol, and running the simulation to its steady state.}
	\label{fig:SERCA_sim}
\end{figure}

Discussion of how bond graph parameters may be determined from an existing kinetic model are presented in Appendix B of the Supplementary Material. The kinetic parameters were thermodynamically constrained, therefore it was possible to find a set of corresponding bond graph parameters by using Eq. B.9 of the Supplementary Material, with the stoichiometric matrix and resulting bond graph parameters given in Appendix D. A comparison of the kinetic and bond graph models (\autoref{fig:SERCA_sim}A) confirms that the two models match closely, with only minor discrepancies at higher cycling rates due the assumption of rapid equilibrium in the \citet{tran_thermodynamic_2009} model.

SERCA accounts for approximately 10--15\% of energy consumption within cardiomyocytes \citep{tran_regulation_2015,schramm_energy_1994}, and is seen as a major contributor to myocardial energy expenditure. In heart failure, SERCA activity decreases, which causes a higher proportion of Ca\textsuperscript{2+} to be removed via an alternative pathway -- the less efficient Na\tsp{}/Ca\textsuperscript{2+} exchanger (NCX) -- increasing energy expenditure \citep{kawase_cardiac_2008}. Therefore it is important to introduce the notion of energy and efficiency into models of transporters in cardiac cells to incorporate the metabolite dependencies required for studying changes in flux under heart failure, and to compare the efficiencies of transporters that move the same substrates. While the parameters for kinetic models can be chosen such that they are thermodynamically consistent, energy-related quantities such as Gibbs free energy and power consumption do not arise naturally from these parameters. Because bond graphs explicitly model energy transfer, an advantage of using this framework is that the power consumption and efficiency are easily calculated from the model. At steady state, the Gibbs free energy of the pump can be calculated from the chemical potentials of metabolites \citep{gawthrop_energy-based_2017}:
\begin{align}
	\Delta G = 2\mu_\mathrm{Casr} + \mu_\mathrm{MgADP} + \mu_\mathrm{Pi} + \mu_\mathrm{H} - 2\mu_\mathrm{Cai} - \mu_\mathrm{MgATP}
\end{align}
The relationship between free energy and SR Ca\textsuperscript{2+} is shown in \autoref{fig:SERCA_sim}B. Note that the Gibbs free energy is zero at the equilibrium point of the pump in \autoref{fig:SERCA_sim}A, as expected of a thermodynamically consistent system. It is important to note that when $\Delta G > 0$, SERCA has been experimentally been observed to operate in the reverse direction where Ca\textsuperscript{2+} flows from the SR to cytosol, and ATP is synthesised \citep{makinose_atp_1971}. We observe in \autoref{fig:SERCA_sim}B that a thermodynamic framework captures this reversal mode of operation, as it is a fundamental physical constraint. The product of free energy and reaction rate gives the rate of power dissipation. As seen in \autoref{fig:SERCA_sim}C, the power consumption is positive under all conditions except at equilibrium, where it is zero.

Because the bond graph approach can split energetic contributions from different sources, it is possible to assess the efficiency of the pump. For this, we define the affinity of ATP hydrolysis as
\begin{align}
A_\text{hyd} = \mu_\mathrm{MgATP} - \mu_\mathrm{MgADP} - \mu_\mathrm{Pi} - \mu_\mathrm{H}
\end{align}
and the affinity of calcium transport as
\begin{align}
A_\text{tr} = 2\mu_\mathrm{Cai} - 2\mu_\mathrm{Casr}
\end{align}
Using the notion of pumping efficiency introduced in \citet{GawCra18}, we define the efficiency $\rho$ as 
\begin{align}
\rho = \begin{cases}
	-A_\text{tr}/A_\text{hyd},&  A_\text{hyd} \ge -A_\text{tr} \\
	-A_\text{hyd}/A_\text{tr},& A_\text{hyd} < -A_\text{tr}
\end{cases}
\end{align}
Thus if the pump operates in the forward direction, efficiency is the proportion of energy from ATP hydrolysis that is converted into energy for calcium transport, and if the pump operates in the reverse direction, efficiency is the proportion of energy supplied from calcium transport that is used to generate ATP. We therefore expect a cusp point at $A_\text{hyd} = -A_\text{tr}$. The efficiency of the pump is plotted in \autoref{fig:SERCA_sim}D. This SERCA model operates reasonably efficiently under the simulated conditions, ranging from 70--100\%, consistent with previous estimates that the pump is 85-90\% efficient \citep{pinz_compromised_2011}. We note also a negative relationship between cycling rate and efficiency, with the pump becoming less efficient the further it moves away from equilibrium. Conversely, the pump approaches an efficiency of 100\% as it nears equilibrium. However, it should be noted that in reality, SERCA pumps may exhibit ``slippage'', where the pump hydrolyses ATP without transporting any Ca\textsuperscript{2+} \citep{meis_ca2+-atpases_2002}. Incorporating such behaviour into the model would likely reduce the maximum operating efficiency of the pump \citep{gawthrop_energy-based_2017}.

\subsection{Na\tsp{}/K\tsp{} ATPase}
\label{sec:NaK_pump}
\begin{figure}
	\centering
	\includegraphics[width=\linewidth]{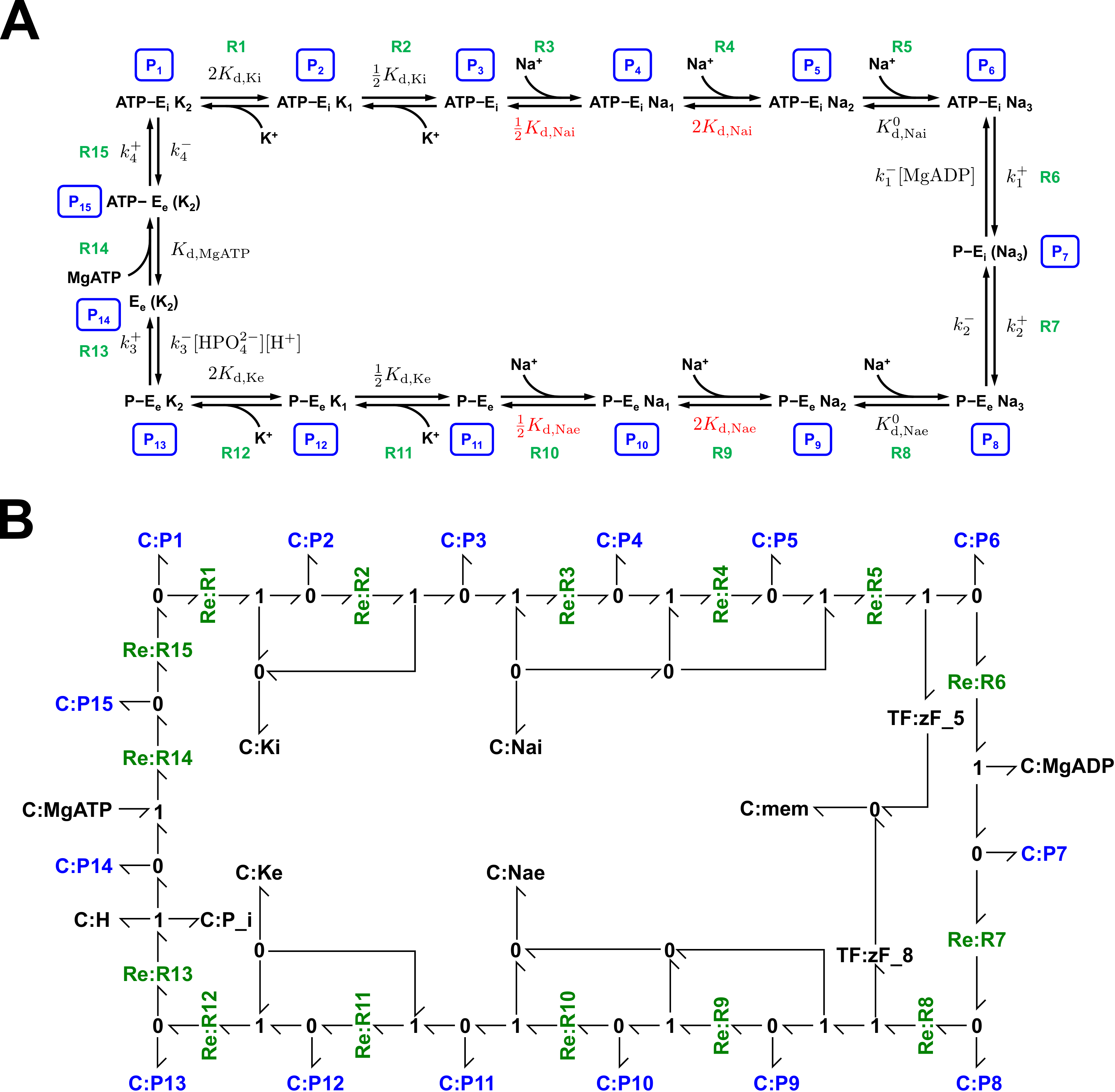}
	\caption{\textbf{The cardiac Na$^+$/K$^+$ ATPase model.} \textbf{(A)} Kinetic model, with numbers for each pump state (\textit{blue boxes}) and reaction names (\textit{green}) are labelled, with corrected parameters shown in red. \textbf{(B)} Bond graph model, with pump states coloured in blue, and reactions coloured in green.}
	\label{fig:NaK_scheme}
\end{figure}
The Na\tsp{}/K\tsp{} ATPase is responsible for maintaining the Na\tsp{} and K\tsp{} gradients that drive ionic currents during the action potential of cardiac cells and many other excitable cells. Na\tsp{} and K\tsp{} are pumped against their electrochemical gradients, therefore their transport requires the supply of energy from ATP hydrolysis. The overall reaction of the pump is
\begin{align}
3\text{Na}_\text{i}^{+} + 2\text{K}_\text{e}^{+} + \text{MgATP} \rightleftharpoons
3\text{Na}_\text{e}^{+} + 2\text{K}_\text{i}^{+} + \text{MgADP} +\text{P}_\text{i} + \text{H}^+ 
\end{align}
In contrast to SERCA, in which only chemical potentials determine the direction of operation, the Na\tsp{}/K\tsp{} ATPase is driven by both chemical and electrical potentials because the plasma membrane is electrically charged. Therefore a thermodynamically consistent model of Na\tsp{}/K\tsp{} ATPase must account for both the free energy of ATP hydrolysis as well as the energetic contribution of the membrane potential.

In this section, we outline a new model of cardiac Na\tsp{}/K\tsp{} ATPase, based on the earlier model by \citet{terkildsen_balance_2007} (\autoref{fig:NaK_scheme}). While the model of \citet{terkildsen_balance_2007} is a biophysically detailed model that incorporates some thermodynamic principles, the final model is thermodynamically inconsistent. We updated the model, correcting the equilibrium constants used for identical binding sites, the equilibrium constant for ATP hydrolysis and mathematical expressions arising from the rapid equilibrium approximation, as described in Appendix C of the Supplementary Material.

\begin{figure}
	\centering
	\includegraphics[width=\linewidth]{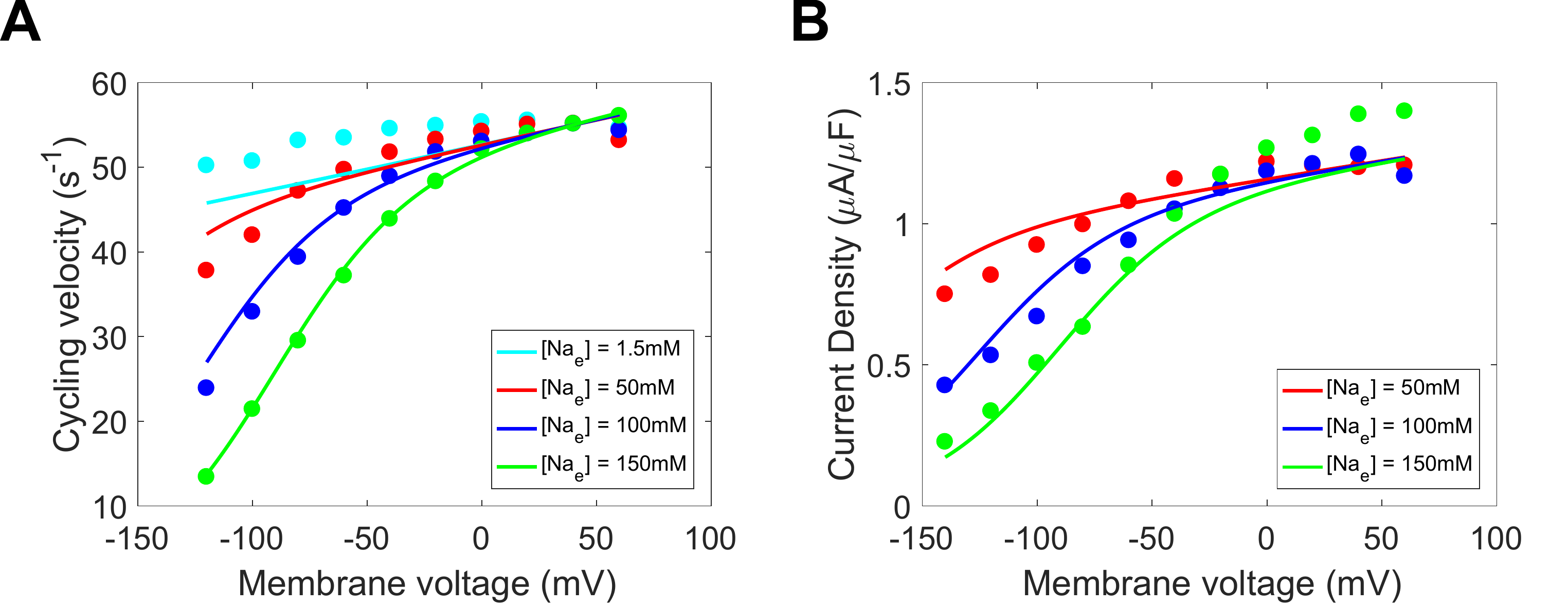}
	\caption{\textbf{Fit of the cardiac Na$^+$/K$^+$ ATPase model to current-voltage measurements.} \textbf{(A)} Comparison of the model to extracellular sodium and voltage data \citep[Fig. 3]{nakao_[na]_1989}, with cycling velocities scaled to a value of $55\si{s^{-1}}$ at $V = 40\si{mV}$. \textbf{(B)} Comparison of the model to whole-cell current measurements \citep[Fig. 2A]{nakao_[na]_1989}.  $\mathrm{[Na^+]_i} = 50\ \si{mM}$, $\mathrm{[K^+]_i} = 0\ \si{mM}$, $\mathrm{[K^+]_e} = 5.4\ \si{mM}$, $\mathrm{pH} = 7.4$, $\mathrm{[Pi]_{tot}} = 0\ \si{mM}$, $\mathrm{[MgATP]} = 10\ \si{mM}$, $\mathrm{[MgADP]} = 0\ \si{mM}$, $T = 310\ \si{K}$.} 
	\label{fig:fitting}
\end{figure}

\begin{figure}
	\centering
	\includegraphics[width=0.8\linewidth]{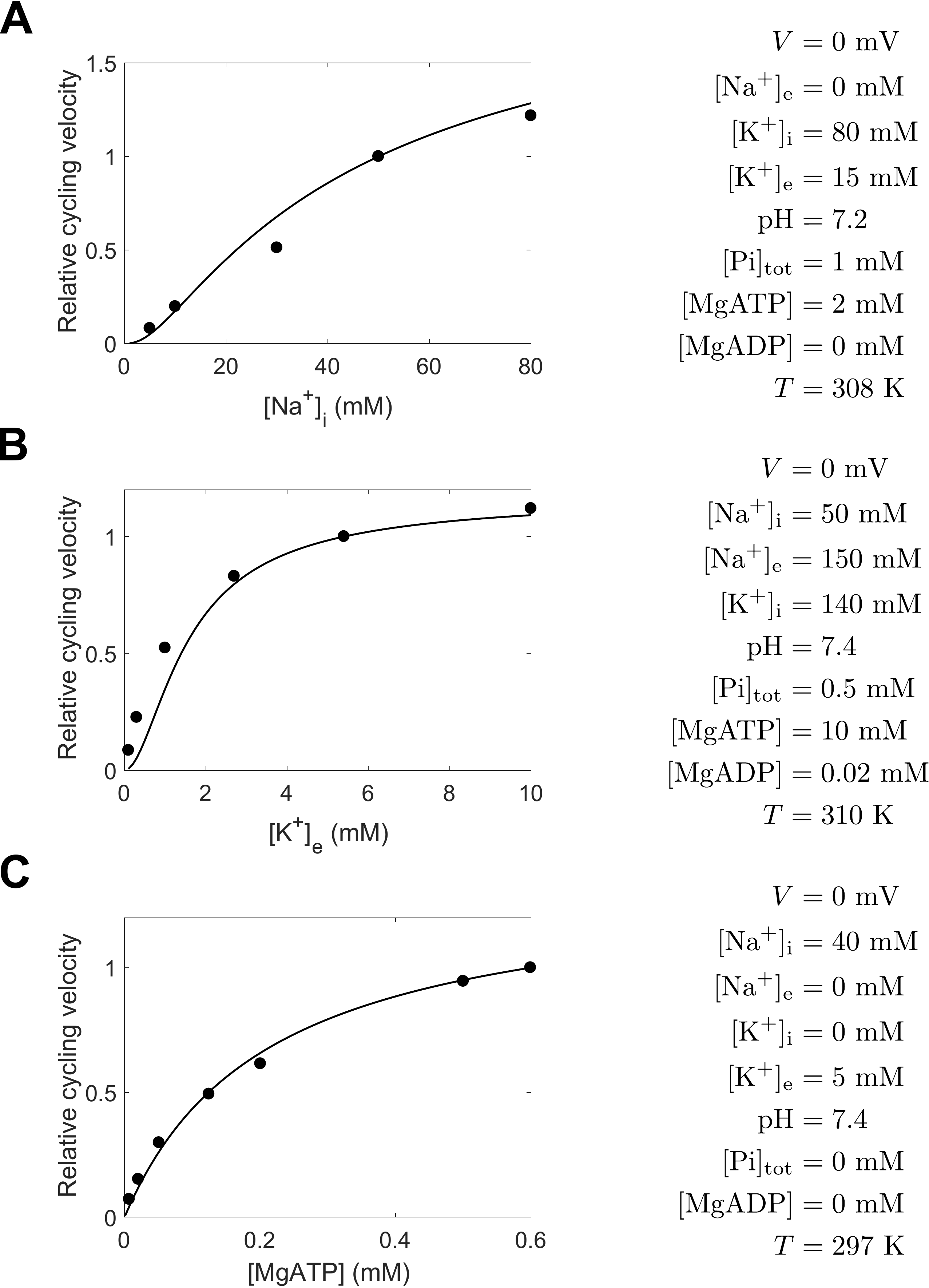}
	\caption{\textbf{Fit of the cardiac Na$^+$/K$^+$ ATPase model to metabolite dependence data.} Simulation conditions are displayed on the right of each figure. \textbf{(A)} Comparison of the model to data with varying intracellular sodium concentrations \citep[Fig. 7A]{hansen_dependence_2002}, normalised to the cycling velocity at $\mathrm{[Na^+]_i = 50\ \si{mM}}$. \textbf{(B)} Comparison of the model to data with varying extracellular potassium \citep[Fig. 11A]{nakao_[na]_1989}, normalised to the cycling velocity at $\mathrm{[K^+]_e = 5.4\ \si{mM}}$. \textbf{(C) } Comparison of the model to data with varying ATP \citep[Fig. 3B]{friedrich_na+k+-atpase_1996}, normalised to the cycling velocity at $\mathrm{[MgATP] = 0.6\ \si{mM}}$.}
	\label{fig:metabolite_dependence}
\end{figure}

We refitted the new model to the data used to parameterise the original \citet{terkildsen_balance_2007} model (see Appendix C of the Supplementary Material for further detail). Comparisons of model simulations to data are given in Figures \ref{fig:fitting},\ref{fig:metabolite_dependence}. The updated model provides a a good fit to the data, and has a comparable fit compared to the original model. The fit in \autoref{fig:fitting}A was slightly worse than the original model at lower extracellular Na\textsuperscript{+} concentrations, but the model seems to be more consistent with experimental evidence that the saturated pump velocity has little sensitivity to extracellular Na\textsuperscript{+} at positive membrane potentials \citep{nakao_[na]_1989}.

\begin{figure}
	\centering
	\includegraphics[width=\linewidth]{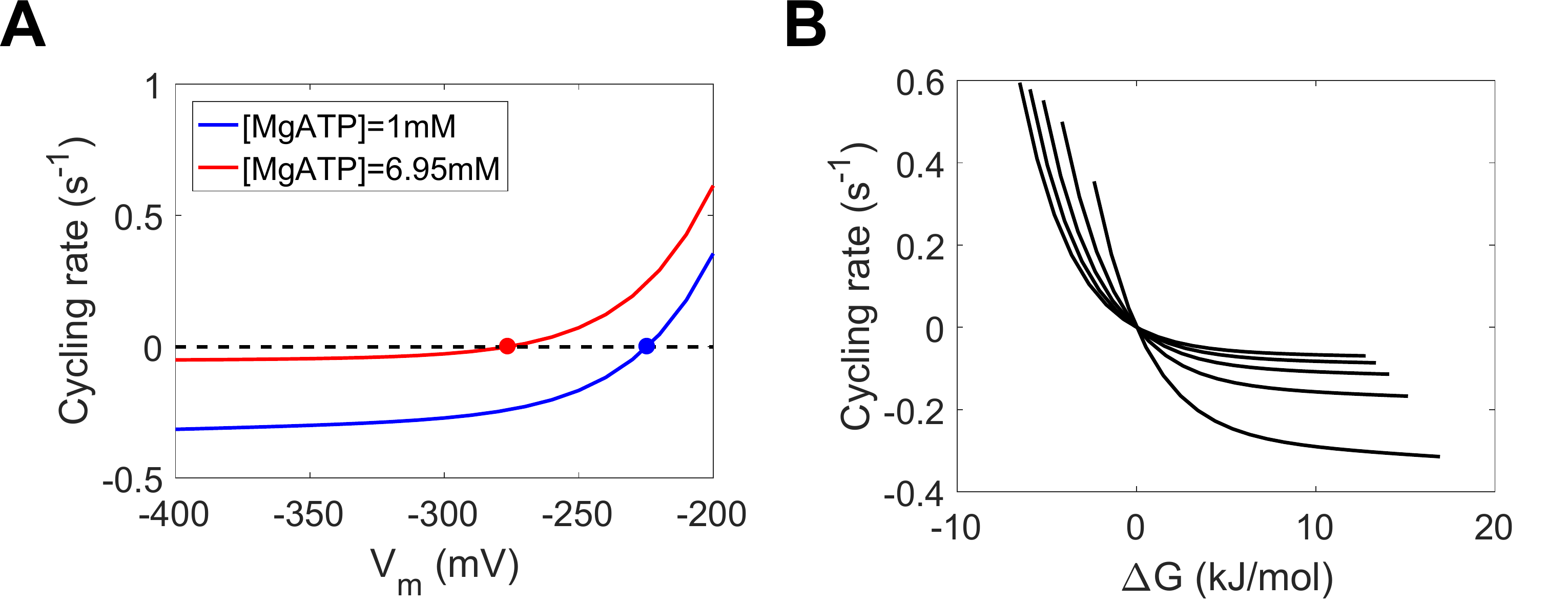}
	\caption{\textbf{Simulation of the Na\tsp{}/K\tsp{} ATPase.} \textbf{(A)} Cycling rates of the pump near reversal potential; \textbf{(B)} Relationship between Gibbs free energy and cycling rate. The curves represent different concentrations of MgATP, from a concentration of 1 mM on the right, with increments of 1 mM up to a concentration of 5mM on the left. The Gibbs free energy was varied by changing the membrane potential. For (A) and (B), simulations were run using $\mathrm{[Na^+]_i} = 10\ \si{mM}$, $\mathrm{[Na^+]_e} = 140\ \si{mM}$, $\mathrm{[K^+]_i} = 145\ \si{mM}$, $\mathrm{[K^+]_e} = 5.4\ \si{mM}$, $\mathrm{pH} = 7.095$, $\mathrm{[Pi]} = 0.3971\ \si{mM}$, $\mathrm{[MgATP]} = 6.95\ \si{mM}$, $\mathrm{[MgADP]} = 0.035\ \si{mM}$. Each pump state was initialised to 1/15 fmol, and steady states were estimated by running each simulation to steady state.}
	\label{fig:NaK_sim}
\end{figure}

The updated model is thermodynamically consistent, therefore it has a bond graph representation (\autoref{fig:NaK_scheme}B). The definition of bond graph parameters and stoichiometric matrices from the fitting process are given in Appendix E of the Supplementary Material. At steady state, the Gibbs free energy of the cycle is
\begin{align}
	\Delta G &= 3\mu_\mathrm{Nae} + 2\mu_\mathrm{Ki} + \mu_\mathrm{MgADP} + \mu_\mathrm{Pi} + \mu_\mathrm{H} - 3\mu_\mathrm{Nai} - 2\mu_\mathrm{Ke} - \mu_\mathrm{MgATP} - \mu_\text{mem} \\
	&= \Delta G_\mathrm{MgATP}^0 + RT\ln \left(
		 \frac{[\mathrm{MgADP}][\mathrm{Pi}][\mathrm{H^+}]}{[\mathrm{MgATP}] } 
	\right)
	+ 3RT\ln \left(
	\frac{[\mathrm{Na^+}]_\text{e}}{[\mathrm{Na^+}]_\text{i} } 
	\right) \nonumber \\
	& \qquad + 2RT\ln \left(
	\frac{[\mathrm{K^+}]_\text{i}}{[\mathrm{K^+}]_\text{e} } 
	\right) - FV_m
\end{align}
To verify the thermodynamic consistency of the bond graph model, we simulated the steady-state operation of the pump at voltages near equilibrium, under physiological (6.95 mM) and ischaemic (1 mM) MgATP concentrations (\autoref{fig:NaK_sim}A). The equilibrium potentials of the pump, where cycling rates are zero, are indicated by dots. An expression for the equilibrium potential can be derived by solving for $\Delta G = 0$:
\begin{align}
	V_m^\text{eq} = \frac{1}{F} \Delta G_\mathrm{MgATP}^0 + \frac{RT}{F} \left(
		\ln 
		\frac{[\mathrm{MgADP}][\mathrm{Pi}][\mathrm{H^+}]}{[\mathrm{MgATP}] } 
		+ 3\ln 
		\frac{[\mathrm{Na^+}]_\text{e}}{[\mathrm{Na^+}]_\text{i} } 
		+ 2\ln 
		\frac{[\mathrm{K^+}]_\text{i}}{[\mathrm{K^+}]_\text{e} }
	\right)
\end{align}

The predicted equilibrium potentials of $-224.5$ mV and $-276.3$ mV for $[\mathrm{MgATP}] = 1\ \si{mM}$ and 6.95 mM respectively are consistent with the equilibria of the steady-state simulations, verifying thermodynamic consistency. We plotted the free energy against cycling rate for a number of MgATP concentrations, and varying the membrane potential to generate the curves in \autoref{fig:NaK_sim}B. Despite differences in the curves, they all satisfy the fundamental constraint that $v_\text{cyc} = 0$ when $\Delta G = 0$, and $v_\text{cyc}$ is positive only at negative free energies. Thus \autoref{fig:NaK_sim} verifies that the updated model is thermodynamically consistent. Like SERCA, the Na\tsp{}/K\tsp{} ATPase is reversible, and has been experimentally observed to synthesise ATP under artificially large ionic gradients \citep{glynn_hundred_2002}. A thermodynamic framework is ideal for describing such phenomena.

\section{Discussion}
In this paper we have discussed the essential thermodynamic principles underlying membrane transporters, and how the bond graph framework captures thermodynamic constraints in these systems. We illustrated using hypothetical models that bond graphs can be used to model simple membrane transport mechanisms, incorporating relevant physical and thermodynamic constraints for both nonelectrogenic and electrogenic transport. The bond graph approach provides a single framework that naturally incorporates known thermodynamic constraints in these systems, including detailed balance \citep{keener_mathematical_2009,liebermeister_modular_2010} and the Nernst potential \citep{keener_mathematical_2009}.

We applied the bond graph approach to two membrane transporters. Using the model of cardiac SERCA by \citet{tran_thermodynamic_2009}, we demonstrated that thermodynamically consistent models can be represented as bond graphs, and that energetic quantities are easily calculated from the bond graph model (\autoref{sec:SERCA}). We then developed a bond graph model of cardiac Na\tsp{}/K\tsp{} ATPase based on earlier work by \citet{terkildsen_balance_2007} and \citet{smith_development_2004}.

\subsection{Thermodynamic constraints}
Because bond graphs are a general-purpose modelling framework for physical systems, they unify a number of known thermodynamic constraints in the literature under a single framework, allowing thermodynamic constraints to be applied more consistently across models of biological systems. Whereas in kinetic models, Wegscheider conditions and detailed balance are used to constrain kinetic parameters in models of membrane transporters \citep{keener_mathematical_2009,smith_development_2004,tran_thermodynamic_2009}, the thermodynamic parameters in a bond graph model automatically account for this constraint \citep{gawthrop_energy-based_2014}. In electrogenic transport, the Nernst equation \citep{keener_mathematical_2009} is a thermodynamic constraint that captures the electrochemical equilibrium. This constraint is automatically accounted for in bond graph modelling, because the Nernst potential is automatically derived from the constitutive equations of the \textbf{C} components for charged species, and the membrane potential \citep{gawthrop_bond_2017-1}.

We believe that thermodynamically consistent models are more likely to remain robust when incorporated into whole-cell models and multi-scale models. It is often uncertain whether models remain predictive under conditions beyond the data they were parameterised on \citep{beattie_sinusoidal_2018,fink_cardiac_2011}. As discussed in \citet{smith_development_2004} and \citet{tran_thermodynamic_2009}, many existing models of transporters do not incorporate dependence for all metabolites and are therefore unable to accurately predict activity under varying metabolite concentrations. Because bond graphs enforce physical constraints on models, they provide useful constraints that dictate the behaviour of the model outside of the range of experimental data \citep{soh_network_2010}, being particularly useful if the experimental data were captured away from equilibrium. In order to satisfy thermodynamic consistency, bond graphs force the modeller to incorporate dependence for all metabolites in models of membrane transporters. Adherence to conservation laws may also prove important when transporter models are incorporated into multi-scale models. In models of cardiac electrophysiology, mass and charge conservation impact on long-term behaviour, and violating these conservation laws may cause the model to drift \citep{hund_ionic_2001,livshitz_uniqueness_2009,pan_bond_2018-1}. Thus in multi-scale models where many individual submodules are simulated over time-scales far greater than they were originally developed for, accounting for conservation laws such as conservation of energy may be an important factor in maintaining the stability of these models.

\subsection{Bond graph models of biological systems}
In this paper, we described the representation of existing models into bond graphs, using a model of SERCA \citep{tran_thermodynamic_2009} as an example. These methods can be applied to a wide range existing thermodynamically consistent transporter models. A bond graph representation reveals the underlying energetics of these models \citep{gawthrop_energy-based_2017}. In addition to membrane transporters, models of ion channels \citep{gawthrop_bond_2017-1}, redox reactions \citep{gawthrop_bond_2017}, metabolic systems \citep{gawthrop_hierarchical_2015} and signalling pathways \citep{gawthrop_modular_2016} can also be represented as bond graphs. As the bond graph methodology continues to develop, we anticipate that more biological processes such as cross-bridge cycling, diffusion and gene regulation will be able to be modelled using bond graphs. Using the modularity inherent in bond graphs \citep{gawthrop_bond_2017}, these models can be coupled together to assemble whole-cell models that can assess how cells deal with fundamental trade-offs between energy efficiency, speed and robustness.

While the bond graph approach can detect thermodynamic inconsistency, it is unable to detect if its thermodynamic parameters are correct. Therefore, in cases such as the model of Na\tsp{}/K\tsp{} ATPase by \citet{terkildsen_balance_2007}, parameterising the model with incorrect species constants or equilibrium constants can lead to incorrect equilibrium points even if the model itself is internally consistent. However, we note that the bond graph framework will flag these issues to the modeller when two models with conflicting thermodynamic parameters are coupled together.

\subsection{Future work}
To date, a large proportion of bond graph models of biochemical systems have been derived from existing kinetic models \citep{gawthrop_hierarchical_2015,gawthrop_bond_2017,gawthrop_energy-based_2017}. However, as outlined in this paper, developing new models using the bond graph framework is a more powerful approach as it guarantees thermodynamic consistency. A challenge in this process is estimating bond graph parameters from data, as kinetic information alone is insufficient to uniquely determine bond graph parameters \citep{gawthrop_hierarchical_2015}. Therefore future work will focus on using existing information and optimisation procedures to derive appropriate estimates of bond graph parameters from data when thermodynamic data is unavailable, and designing efficient experimental protocols for identifying bond graph parameters when thermodynamic data is available. In this process, there is a key role for quantifying the uncertainty of biological parameters \citep{babtie_how_2017,beattie_sinusoidal_2018}. Given that bond graph parameters are physical quantities, it would be interesting to investigate whether these uncertainties propagate as individual models are coupled together.

We modelled transporters using multiple states with mass action equations, giving rise to steady-state velocities. However, many models of transporters are simplifications of this approach, as they are described by a single equation for the transport rate rather than multiple equations for transitions between enzyme states \citep{keener_mathematical_2009}. \citet{smith_development_2004} describe a method for reducing multi-state models of membrane transporters into a single equation for transport rate by using rapid equilibrium and steady-state simplifications, while using detailed balance to ensure thermodynamic consistency. As we move towards increasingly complex multi-scale models, it is important to develop methods of simplifying models to reduce computational cost \citep{smith_computational_2007}. An advantage of using bond graphs in the simplification process is that they impose constraints to ensure that simplified models maintain thermodynamic consistency \citep{gawthrop_energy-based_2014}.

\section{Conclusion}
We have shown that bond graphs can be used to model membrane transporters while capturing fundamental thermodynamic and physical constraints. We apply this framework to SERCA and Na\tsp{}/K\tsp{} ATPase to develop models that are thermodynamically consistent while revealing the underlying energetics of these transporters. Combined with their inherent modularity, we believe that bond graphs are a powerful tool for incorporating models of membrane transporters into other models while maintaining thermodynamic consistency.

\section*{Data availability}
The code associated with this paper is available from GitHub (\url{https://github.com/uomsystemsbiology/transporter_thermodynamics}), and is archived on Zenodo (\url{https://doi.org/10.5281/zenodo.1287353}) \citep{pan_supporting_2018-1}. The repositories contain MATLAB code that generates the figures and CellML code containing equations for the models.

\section*{Acknowledgements}
This research was in part conducted and funded by the Australian Research Council Centre of Excellence in Convergent Bio-Nano Science and Technology (project number CE140100036), and the Australian Research Council's Discovery Projects funding scheme (project DP170101358). M.P. would like to acknowledge financial support provided by an Australian Government Research Training Program Scholarship. P.J.G. would like to thank the Melbourne School of Engineering for its support via a Professorial Fellowship. K.T. is supported by the Heart Foundation of New Zealand (Research Fellowship 1692) and the Marsden Fund Council from Government funding, managed by Royal Society Te Apārangi (Marsden Fast-Start UOA1703).

\bibliographystyle{elsarticle-harv}
\small
\bibliography{bibliography,bibliography2}{}
\normalsize

\newpage
\appendix
\setcounter{figure}{0}
\setcounter{table}{0}

\section{TF components in biochemical networks}
\label{sec:TF_biochemical}
\begin{figure}[b]
	\centering
	\includegraphics[width=0.45\linewidth]{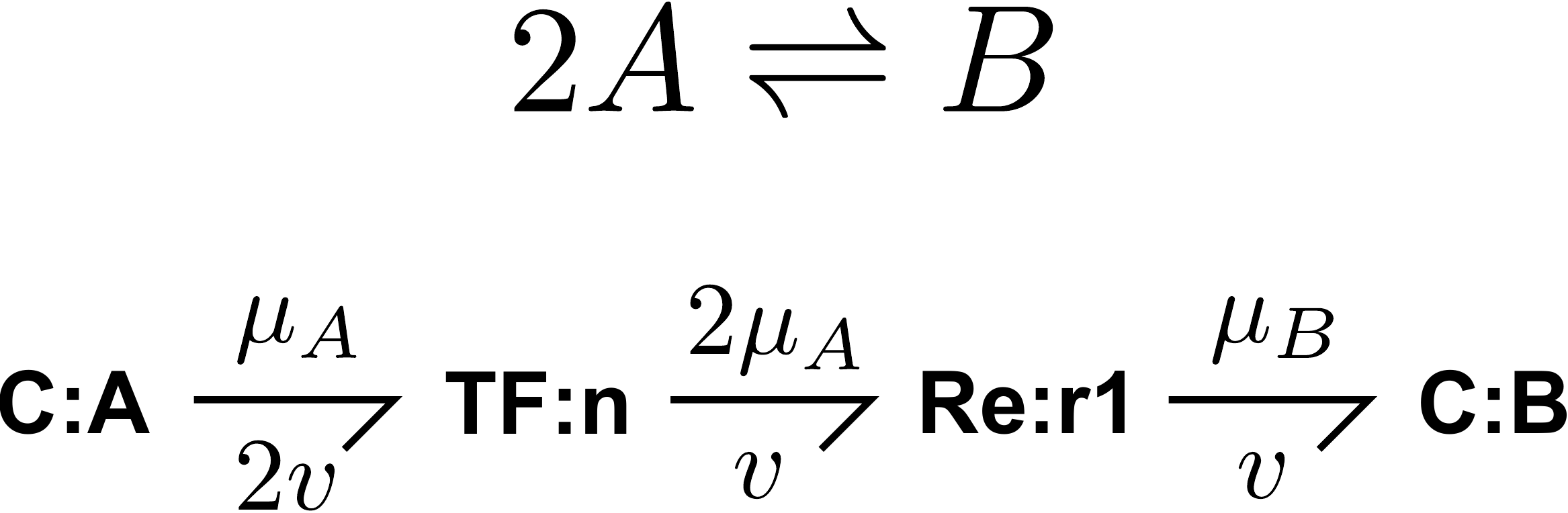}
	\caption{\textbf{Bond graph of the reaction $A \rlh B$.}}
	\label{fig:TF_biochemical}
\end{figure}

For some reactions such as dimerisation, two or more of the same species may be involved in one side of the reaction. To represent this, the bond connected to that species is fed through a \textbf{TF} component. The \textbf{TF} component represents a transformer, which can transmit and convert energy while maintaining energy conservation. The equation for the transformer is
\begin{align}
	e_2 &= n e_1 \\
	f_1 &= n f_2
\end{align}
Where $n$ represents the stoichiometry of the species. Since $e_1f_1 = e_2 f_2$, the transformer is energy-conserving as the power in is equal to the power out of the transformer.

When applied the biochemical reaction $2A \rightleftharpoons B$, with the bond graph representation in \autoref{fig:TF_biochemical}, the constitutive equations of the \textbf{Re} and \textbf{C} components are
\begin{align}
	\mu_A &= RT\ln (K_A x_A) \\
	\mu_B &= RT\ln (K_B x_B) \\
	v &= \kappa (e^{A^f / RT} - e^{A^r / RT}) 
\end{align}
The equation corresponding to the transformer (with $n=2$) is
\begin{align}
	A_1^f &= 2\mu_A \\
	\dot{x}_A &= -2v
\end{align}
thus the transformer states that the rate of consumption of A is twice the rate of reaction. Because two molecules of A are consumed by the reaction, the forward affinity is twice the chemical potential of A. By combining these equations the law of mass action can be derived:
\begin{align}
	v &= \kappa (e^{2\mu_A / RT} - e^{\mu_B / RT}) = \kappa K_A^2 x_A^2 - \kappa K_B x_B \\
	\dot{x}_A &= -2v = 2\kappa K_B x_B -2\kappa_1 K_A^2 x_A^2 \\
	\dot{x}_B &= v = \kappa K_A^2 x_A^2 - \kappa K_B x_B
\end{align}

\section{Converting existing kinetic models to bond graphs}
\label{sec:conversion}
Because there are a number of existing models of membrane transporters that are thermodynamically consistent, the ability to convert kinetic models of transporters into bond graphs would save a great amount of effort from having to build new bond graph models to explain existing data. In \autoref{sec:enzyme_cycle}, we outlined how a chemical reaction network can be represented as a bond graph. While the structure of the system can be naturally translated a bond graph, the parameters of existing models do not directly translate into bond graph parameters $K$ and $\kappa$. While bond graph parameters define a unique set of kinetic parameters, identifying bond graph parameters from a set of kinetic parameters is more involved. In this section, we outline how to infer bond graph parameters from kinetic parameters, and discuss the issues that arise during this process.

We first use the enzyme cycle example in \autoref{sec:enzyme_cycle} to illustrate the essential concepts of inferring bond graph parameters. The relationship between the kinetic and bond graph parameters is given by Eqs. \ref{eq:kinetic_bg_enzyme_start}--\ref{eq:kinetic_bg_enzyme}. Additional constraints are enforced by the detailed balance constraint in Eq. \ref{eq:detailed_balance_simple}. Therefore, given a set of kinetic constants $k_1^+$, $k_2^+$, $k_1^-$, $k_2^-$, an equivalent set of bond graph parameters can be derived by solving the Eqs. \ref{eq:kinetic_bg_enzyme_start}--\ref{eq:kinetic_bg_enzyme}, \ref{eq:detailed_balance_simple}. If we define $\Ln$ as the element-wise logarithm operator, these constraints can be packaged into the matrix equation
\begin{align}
	\Ln \begin{bmatrix}
		k_1^+ \\ k_2^+ \\ k_1^- \\ k_2^- \\ 1
	\end{bmatrix}
	= \begin{bmatrix}
		1 & 0 & 1 & 0 & 1 & 0 \\
		0 & 1 & 0 & 0 & 0 & 1 \\
		1 & 0 & 0 & 0 & 0 & 1 \\
		0 & 1 & 0 & 1 & 1 & 0 \\
		0 & 0 & 1 & -1 & 0 & 0 
	\end{bmatrix}
	\Ln
	\begin{bmatrix}
		\kappa_1 \\ \kappa_2 \\ K_\mathrm{Si} \\ K_\mathrm{Se} \\ K_{E1} \\ K_{E2}
	\end{bmatrix}
	\label{eq:enzyme_matrix}
\end{align}
which can be solved through linear methods \citep{anton_elementary_2014}. However, it should be noted that the rows of the right hand side are linearly dependent, as $R1+R2-R3-R4-R5 = 0$. Therefore, in order for a solution to exist, the kinetic parameters must satisfy the same dependency between rows:
\begin{align}
	\begin{bmatrix}
		1 & 1 & -1 & -1 & -1
	\end{bmatrix}
	\Ln \begin{bmatrix}
		k_1^+ & k_2^+ & k_1^- & k_2^- & 1
	\end{bmatrix}^T = 0
\end{align}
which is easily seen to be the detailed balance constraint in Eq. \ref{eq:detailed_balance_kinetic_enzyme}. As a result, a set of bond graph parameters exist for this kinetic system only if the kinetic parameters are thermodynamically consistent, reflecting the fact that bond graphs can only represent the subset of kinetic systems that are thermodynamically consistent.

The equations shown above can be generalised to all chemical reaction networks described by mass action. The relationships between kinetic and bond graph parameters are captured in the general equation \citep{gawthrop_hierarchical_2015}
\begin{align}
	\textbf{Ln}(\mathbf{k}) = \mathbf{M} \textbf{Ln}(\mathbf{W} \boldsymbol{\lambda}) \label{eq:bg_general}
\end{align}
with partitions defined as
\begin{align}
	\textbf{k} = \begin{bmatrix}
		k^+ \\ k^- \\ K^c
	\end{bmatrix}, \quad
	\textbf{M} = \left[ \begin{array}{c | c}
		I_{n_r \times n_r} & {N^f}^T \\ \hline
		I_{n_r \times n_r} & {N^r}^T \\ \hline
		0 & {N^c}^T
	\end{array} \right], \quad
	\boldsymbol{\lambda} = \begin{bmatrix}
		\kappa \\ K
	\end{bmatrix}
	\label{eq:standard}
\end{align}
Here, $k^+$ and $k^-$ are vectors of the forward and reverse kinetic rate constants respectively, $I_{n \times n}$ is an identity matrix of length $n$, $n_r$ is the number of reactions, $N^f$ and $N^r$ are the forward and reverse stoichiometric matrices respectively, $\kappa$ is a vector of the reaction rate constants, $K$ is a vector of the species thermodynamic constants, $K^c$ is a vector of known equilibrium constants between the species defined in the matrix $N^c$, and $\mathbf{W}$ is a diagonal matrix that accounts for differences in volume between compartments. Therefore the first two rows represent the relationships between kinetic and bond graph constants, and the final row represents additional information that is known about the thermodynamic constants.

For the enzyme cycle, Eq. \ref{eq:bg_general} gives rise to Eq. \ref{eq:enzyme_matrix} through the substitutions
\begin{align}
	&k^+ = \begin{bmatrix}
		k_1^+ \\ k_2^+
	\end{bmatrix} , \qquad
	k^- = \begin{bmatrix}
		k_1^- \\ k_2^-
	\end{bmatrix} , \qquad
	\kappa = \begin{bmatrix}
		\kappa_1 \\ \kappa_2
	\end{bmatrix} , \qquad
	K = \begin{bmatrix}
		K_\mathrm{Si} \\ K_\mathrm{Se} \\ K_{E1} \\ K_{E2}
	\end{bmatrix} , \qquad
	K^c = 1 \\
	&N^f = \begin{bmatrix}
		1 & 0 \\
		0 & 0 \\
		1 & 0 \\
		0 & 1
	\end{bmatrix} , \qquad
	N^r = \begin{bmatrix}
		0 & 0 \\
		0 & 1 \\
		0 & 1 \\
		1 & 0
	\end{bmatrix} , \qquad
	N^c = \begin{bmatrix}
		1 \\ -1 \\ 0 \\ 0
	\end{bmatrix}, \qquad
	\mathbf{W} = I_{6 \times 6}
\end{align}

As seen in the enzyme cycle example, Eq. \ref{eq:bg_general} is not always solvable because there may exist sets of kinetic rate constants that do not satisfy detailed balance constraints, and are therefore thermodynamically inconsistent. Because the bond graph approach forces the modeller to use extra discipline in accounting for energy transfer, the exercise of converting kinetic parameters into bond graphs can be used to verify the thermodynamic consistency of a kinetic model. We show here that bond graph parameters can be derived from a set of kinetic parameters if they satisfy a thermodynamic constraint, known in the literature as Wegscheider conditions \citep{liebermeister_modular_2010}. By dividing the first $n_r$ rows of Eq. \ref{eq:bg_general} by the next $n_r$ rows, we find that
\begin{align}
	\Ln \begin{bmatrix}
		K^\text{eq} \\ K^c
	\end{bmatrix}
	=
	\begin{bmatrix}
		-N^T \\ {N^c}^T
	\end{bmatrix}
	\Ln (K) \label{eq:Keq}
\end{align}
where $N = N^r - N^f$ is the stoichiometric matrix. If we define $/$ as the element wise quotient operator for vectors, $K^\text{eq} = k^+/k^-$ is a vector containing the equilibrium constants of the reactions. Thus if $Z$ is a right nullspace matrix \citep{anton_elementary_2014} of $[-N \quad N^c]$, multiplying both sides of Eq. \ref{eq:Keq} by $Z^T$ on the left gives
\begin{align}
	Z^T \Ln \begin{bmatrix}
		K^\text{eq} \\ K^c
	\end{bmatrix} = 0
	\label{eq:TC_check}
\end{align}
Thus for any biochemical system with mass action equations, Eq. \ref{eq:TC_check} can be used to check that the kinetic parameters are thermodynamically consistent. As discussed in \citet{gawthrop_hierarchical_2015}, Eq. \ref{eq:TC_check} is a form of the Wegscheider condition, which has been associated with thermodynamic consistency \citep{polettini_irreversible_2014}, therefore only thermodynamically consistent kinetic parameters can be converted into bond graph parameters.

Assuming that a solution exists, one solution to Eq. \ref{eq:bg_general} can be found using the equation
\begin{align}
	\boldsymbol{\lambda_0 } = \mathbf{W}^{-1} \textbf{Exp} (\mathbf{M}^\dagger \Ln (\mathbf{k}))
	\label{eq:bg_param_sol}
\end{align}
where $\mathbf{M}^\dagger$ is the Moore-Penrose pseudoinverse of $\mathbf{M}$ \citep{hung_moore-penrose_1975}. If a solution exists, the set of bond graph parameters is generally not unique. However, as demonstrated in \citet{gawthrop_hierarchical_2015}, the affinity and velocity each reaction remains the same regardless of the parameters chosen.

To account for differences in volume between compartments, we include a matrix $\mathbf{W}$ in Eq. \ref{eq:bg_general}. If a species exists within a compartment, the diagonal entry corresponding to that species is equal to the volume of that compartment. For reaction rate constants and species that do not exist within a volume, the value of the corresponding diagonal element is 1.

In \autoref{sec:results}, we use the constraints
\begin{align}
	\frac{W_\text{i} K_\text{Nai}}{W_\text{e} K_\text{Nae}} &= 1 \\
	\frac{W_\text{i} K_\text{Ki}}{W_\text{e} K_\text{Ke}} &= 1 \\
	\frac{W_\text{i} K_\text{Cai}}{W_\text{sr} K_\text{Casr}} &= 1 \\
	\frac{W_\text{i}K_\mathrm{MgATP}}{W_\text{i}K_\mathrm{MgADP} W_\text{i}K_\mathrm{Pi}W_\text{isr}K_\mathrm{H}} &= \exp \left(-\frac{\Delta G_{\text{MgATP}}^0}{RT} \right)\ \si{M}^2 = 9881\ \si{mM^2}
\end{align}
where $\Delta G_{\text{MgATP}}^0 = 11.9 \ \si{kJ/mol}$ is the standard free energy of MgATP hydrolysis at 310~K \citep{guynn_equilibrium_1973,tran_thermodynamic_2009}. We used the volumes $W_\text{i} = 38.0\ \si{pL}$, $W_\text{e} = 5.182\ \si{pL}$, and $W_\text{sr} = 2.28\ \si{pL}$ \citep{luo_dynamic_1994}. For the ATP hydrolysis reaction, we used the intracellular volume for MgATP, MgADP and Pi. In the \citet{terkildsen_balance_2007} model, a volume of $W_\text{isr}=W_\text{i}$ was used for H\tsp{}. The \citet{tran_thermodynamic_2009} model describes the countertransport of Ca\textsuperscript{2+} and H\tsp{}, however the steady-state equations were derived assuming that the cytsol and SR shared the same H\tsp{} concentration. Therefore we chose to use the volume $W_\text{isr}=W_\text{i}+W_\text{sr}$.

In examples where charged species are involved (such as in \autoref{sec:charged_example}), the electrical dependence can be ignored for the purpose of identifying bond graph parameters, since the components corresponding to the electrical dependence (the linear \textbf{C} component, and \textbf{TF} components) have a natural bond graph representation.

The \citet{tran_thermodynamic_2009} and \citet{terkildsen_balance_2007} models assumed that some reactions were at rapid equilibrium, therefore we approximated them in the bond graph model by replacing the equilibrium constants with two sufficiently fast kinetic rate constants with the same equilibrium constant. 

\section{Modifications to the Terkildsen et al. model of the Na\tsp{}/K\tsp{} ATPase}
\label{sec:Terkildsen_modifications}
\subsection{Updates to equations}
The Terkildsen \textit{et al.} model is a 15-state model of the Na\tsp{}/K\tsp{} ATPase (\autoref{fig:NaK_scheme}). The model was reduced to a 4-state model using rapid equilibrium \citep{smith_development_2004}, and then further simplified using steady state approximations. We found kinetic and thermodynamic issues in the implementation of this model, and resolved these issues as follows:
\begin{enumerate}
	\item The equilibrium constants of the chemical reaction network were inconsistent with the number of binding sites. If we assume that the binding/unbinding events are identical, the kinetic rate constants are proportional to the number of binding sites \citep{keener_mathematical_2009}. We have corrected the equilibrium constants, shown in red in \autoref{fig:NaK_scheme}A.
	\item The original model by \citet{terkildsen_balance_2007} used an incorrect standard free energy of $\Delta G_\mathrm{MgATP}^0 = -30.2\ \si{kJ/mol}$ for the hydrolysis of ATP, which resulted in an incorrect equilibrium constant. The authors appeared to adjust $\Delta G_\mathrm{MgATP}^0$ to a physiological pH rather than a pH of 0. Because $\Delta G_\mathrm{MgATP}^0$ was used in the detailed balance constraint
	\begin{align}
		\frac{k_1^+ k_2^+ k_3^+ k_4^+ K_{d,\text{Na}_e}^0(K_{d,\text{Na}_e})^2 (K_{d,\text{K}_i})^2}
		{k_1^- k_2^- k_3^- k_4^- K_{d,\text{Na}_i}^0 (K_{d,\text{Na}_i})^2 (K_{d,\text{K}_e})^2 K_{d,\text{MgATP}}} = \exp\left( -\frac{\Delta G_\text{MgATP}^0}{RT} \right),
		\label{eq:detailed_balance}
	\end{align}
	the model was thermodynamically inconsistent. This error would cause over a $10^7$--fold increase in the equilibrium constant at 310 K, therefore we corrected the value to $\Delta G_\mathrm{MgATP}^0 = 11.9\ \si{kJ/mol}$ \citep{guynn_equilibrium_1973,tran_thermodynamic_2009}. 
	\item \citet{terkildsen_balance_2007} applied a rapid equilibrium approximation to reduce the 15-state model into a 4-state model with modified kinetic constants that were functions of metabolite concentrations. However, due to algebraic errors, the equations for some of these modified kinetic rate constants ($\alpha_1^+$, $\alpha_3^+$, $\alpha_2^-$ and $\alpha_4^-$) were incorrect. In our updated model, we have corrected the following modified rate constants to the equations below:
	\begin{align}
		\alpha_1^+ &= \frac{k_1^+  \tilde{\text{Na}}_{i,1}\tilde{\text{Na}}_{i,2}^2}{\tilde{\text{Na}}_{i,1}\tilde{\text{Na}}_{i,2}^2 + (1 + \tilde{\text{Na}}_{i,2})^2 + (1 + \tilde{\text{K}}_i)^2 -1} \\
		\alpha_3^+ &= \frac{k_3^+ \tilde{\text{K}}_e^{2}}{\tilde{\text{Na}}_{e,1}\tilde{\text{Na}}_{e,2}^2 + (1 + \tilde{\text{Na}}_{e,2})^2 + (1 + \tilde{\text{K}}_e)^2 -1 } \\
		\alpha_2^-&= \frac{k_2^- \tilde{\text{Na}}_{e,1}\tilde{\text{Na}}_{e,2}^2}{\tilde{\text{Na}}_{e,1}\tilde{\text{Na}}_{e,2}^2 + (1 + \tilde{\text{Na}}_{e,2})^2 + (1 + \tilde{\text{K}}_e)^2 -1} \\
		\alpha_4^- &= \frac{k_4^- \tilde{\text{K}}_i^2}{\tilde{\text{Na}}_{i,1}\tilde{\text{Na}}_{i,2}^2 + (1 + \tilde{\text{Na}}_{i,2})^2 + (1 + \tilde{\text{K}}_i)^2 -1}
	\end{align}
	where
	\begin{align}
		&\tilde{\text{Na}}_{i,1} = \frac{\mathrm{[Na^+]_i}}{K_{d,\text{Na}_i}^0 e^{\Delta FV_m/RT}} 
		&&\tilde{\text{Na}}_{i,2} = \frac{\mathrm{[Na^+]_i}}{K_{d,\text{Na}_i}} \\
		&\tilde{\text{Na}}_{e,1} = \frac{\mathrm{[Na^+]_e}}{K_{d,\text{Na}_e^0} e^{(1+\Delta) zFV_m/RT}} 
		&&\tilde{\text{Na}}_{e,2} = \frac{\mathrm{[Na^+]_e}}{K_{d,\text{Na}_e}} \\
		&\tilde{\text{K}}_i = \frac{\mathrm{[K^+]_i}}{K_{d,\text{K}_i}} 
		&&\tilde{\text{K}}_e = \frac{\mathrm{[K^+]_e}}{K_{d,\text{K}_e}} 
	\end{align}
	and $\Delta$ is the charge translocated by reaction R5.
	
	$\alpha_1^+$ was one of the modified rate constants that was updated, described in \citet{terkildsen_balance_2007} by the incorrect equation
	\begin{align}
		\alpha_1^+ =  \frac{k_1^+  \tilde{\text{Na}}_{i,1}\tilde{\text{Na}}_{i,2}^2}{ (1 + \tilde{\text{Na}}_{i,1})(1 + \tilde{\text{Na}}_{i,2})^2 + (1 + \tilde{\text{K}}_i)^2 -1}
	\end{align}
	Since the modified rate constant is $k_1^+$ scaled by the proportion of state 6 with respect to the total amount of states 1 to 6, a correct expression for the modified rate constant can be derived as follows:
	\begin{align}
		\alpha_1^+ &= k_1^+ \frac{x_6}{x_6 + x_5 + x_4 + x_3 + x_2 + x_1} \notag \\
		&= k_1^+ \frac{1}{1 + x_5/x_6 + x_4/x_6 + x_3/x_6 + x_2/x_6 + x_1/x_6} \notag  \\
		&= \frac{k_1^+}{1 + 2\tilde{\text{Na}}_{i,1}^{-1} + 2\tilde{\text{Na}}_{i,1}^{-1}\tilde{\text{Na}}_{i,2}^{-1} + \tilde{\text{Na}}_{i,1}^{-1}\tilde{\text{Na}}_{i,2}^{-2} + 2\tilde{\text{Na}}_{i,1}^{-1}\tilde{\text{Na}}_{i,2}^{-2}\tilde{\text{K}}_i + \tilde{\text{Na}}_{i,1}^{-1}\tilde{\text{Na}}_{i,2}^{-2}\tilde{\text{K}}_i^2} \notag  \\
		&= \frac{k_1^+  \tilde{\text{Na}}_{i,1}\tilde{\text{Na}}_{i,2}^2}{\tilde{\text{Na}}_{i,1}\tilde{\text{Na}}_{i,2}^2 + (1 + \tilde{\text{Na}}_{i,2})^2 + (1 + \tilde{\text{K}}_i)^2 -1}
	\end{align}
	The mathematical expressions for the other modified rate constants can be derived using a similar procedure.
\end{enumerate}

\subsection{Reparameterising the Terkildsen et al. model}
Using the updated equations of the Terkildsen \textit{et al.} model, and setting $\Delta G_\mathrm{MgATP}^0 = 11.9\ \si{kJ/mol}$ for the detailed balance constraint, we reparameterised the model to the same data that the Terkildsen \textit{et al.} model was trained on. We minimised an objective function that summarised differences between model predictions and data, using similar methods to those in \citet{terkildsen_modelling_2006}, with some minor changes:
\begin{enumerate}
	\item For the extracellular K\tsp{} data of \citet{nakao_[na]_1989} (\autoref{fig:metabolite_dependence}B), the weighting for extracellular potassium above 5.4 mM was increased from $6\times$ to $15\times$ to obtain a reasonable fit at physiological concentrations.
	\item To ensure that the magnitude of cycling velocity matched that of \citet{nakao_[na]_1989}, we chose to fit the curve for $\mathrm{[Na^+]_\text{e}} = 150\ \si{mM}$ (\autoref{fig:fitting}A) without normalisation.
	\item Instead of using a local optimiser to minimise the objective function, we used particle swarm optimisation \citep{kennedy_particle_1995} followed by a local optimiser to find a global minimum.
\end{enumerate}

\section{Stoichiometry and parameters for SERCA}
\label{sec:SERCA_params}
\begin{table}[H]
	\centering
	\caption{\textbf{Forward stoichiometric matrix for the \citet{tran_thermodynamic_2009} SERCA model.}}
	\label{tab:SERCA_Nf}
	\begin{tabular}{cccccccccc}
		\toprule
		& R1$\to$2   & R2$\to$4 & R2$\to$2a & R4$\to$5 & R5$\to$6 & R6$\to$8 & R8$\to$9 & R9$\to$10 & R10$\to$1     \\ \midrule
		P1     & 1    & 0     & 0    & 0    & 0    & 0    & 0     & 0     & 0 \\
		P2     & 0    & 1     & 1    & 0    & 0    & 0    & 0     & 0     & 0 \\
		P2a    & 0    & 0     & 0    & 0    & 0    & 0    & 0     & 0     & 0 \\
		P4     & 0    & 0     & 0    & 1    & 0    & 0    & 0     & 0     & 0 \\
		P5     & 0    & 0     & 0    & 0    & 1    & 0    & 0     & 0     & 0 \\
		P6     & 0    & 0     & 0    & 0    & 0    & 1    & 0     & 0     & 0 \\
		P8     & 0    & 0     & 0    & 0    & 0    & 0    & 1     & 0     & 0 \\
		P9     & 0    & 0     & 0    & 0    & 0    & 0    & 0     & 1     & 0 \\
		P10    & 0    & 0     & 0    & 0    & 0    & 0    & 0     & 0     & 1 \\
		H\tsp{}     & 0    & 0     & 1    & 0    & 0    & 0    & 2     & 0     & 0 \\
		$\mathrm{Ca^{2+}_i}$  & 0    & 2     & 0    & 0    & 0    & 0    & 0     & 0     & 0 \\
		$\mathrm{Ca^{2+}_{sr}}$ & 0    & 0     & 0    & 0    & 0    & 0    & 0     & 0     & 0 \\
		MgATP  & 1    & 0     & 0    & 0    & 0    & 0    & 0     & 0     & 0 \\
		MgADP  & 0    & 0     & 0    & 0    & 0    & 0    & 0     & 0     & 0 \\
		Pi     & 0    & 0     & 0    & 0    & 0    & 0    & 0     & 0     & 0 \\ \bottomrule
	\end{tabular}
\end{table}

\begin{table}[H]
	\centering
	\caption{\textbf{Reverse stoichiometric matrix for the \citet{tran_thermodynamic_2009} SERCA model.}}
	\label{tab:SERCA_Nr}
	\begin{tabular}{cccccccccc}
		\toprule
		& R1$\to$2   & R2$\to$4 & R2$\to$2a & R4$\to$5 & R5$\to$6 & R6$\to$8 & R8$\to$9 & R9$\to$10 & R10$\to$1     \\ \midrule
		P1     & 0    & 0     & 0    & 0    & 0    & 0    & 0     & 0     & 1 \\
		P2     & 1    & 0     & 0    & 0    & 0    & 0    & 0     & 0     & 0 \\
		P2a    & 0    & 0     & 1    & 0    & 0    & 0    & 0     & 0     & 0 \\
		P4     & 0    & 1     & 0    & 0    & 0    & 0    & 0     & 0     & 0 \\
		P5     & 0    & 0     & 0    & 1    & 0    & 0    & 0     & 0     & 0 \\
		P6     & 0    & 0     & 0    & 0    & 1    & 0    & 0     & 0     & 0 \\
		P8     & 0    & 0     & 0    & 0    & 0    & 1    & 0     & 0     & 0 \\
		P9     & 0    & 0     & 0    & 0    & 0    & 0    & 1     & 0     & 0 \\
		P10    & 0    & 0     & 0    & 0    & 0    & 0    & 0     & 1     & 0 \\
		H\tsp{}     & 0    & 0     & 0    & 2    & 0    & 0    & 0     & 1     & 0 \\
		$\mathrm{Ca^{2+}_i}$  & 0    & 0     & 0    & 0    & 0    & 0    & 0     & 0     & 0 \\
		$\mathrm{Ca^{2+}_{sr}}$ & 0    & 0     & 0    & 0    & 0    & 2    & 0     & 0     & 0 \\
		MgATP  & 0    & 0     & 0    & 0    & 0    & 0    & 0     & 0     & 0 \\
		MgADP  & 0    & 0     & 0    & 0    & 1    & 0    & 0     & 0     & 0 \\
		Pi     & 0    & 0     & 0    & 0    & 0    & 0    & 0     & 0     & 1 \\ \bottomrule
	\end{tabular}
\end{table}

\begin{table}[H]
	\caption{\textbf{Parameters for the Tran SERCA model.} Adapted from \citet{tran_thermodynamic_2009}.}
	\centering
	\begin{tabular}{l l l}
		\toprule
		\textbf{Parameter} & \textbf{Value} & \textbf{Modified kinetic constants}\\ \midrule
		$k_1^+$ & $25900 \si{mM^{-1} s^{-1}}$ & - \\
		$k_2^+$ & $2540 \si{s^{-1}}$& - \\
		$k_3^+$ & $20.5 \si{s^{-1}}$& - \\
		$k_1^-$ & $2 \si{s^{-1}}$& - \\
		$k_2^-$ & $67200 \si{mM^{-1}s^{-1}}$& - \\
		$k_3^+$ & $149 \si{mM^{-1}s^{-1}}$& - \\
		$K_\text{d,Cai}^\text{eq}$ & 0.91mM & $k^+ = 3.1276\times 10^{10}\si{mM^{-2} s^{-1}}$\\ & & $k^- = 2.5900\times 10^{10}\si{s^{-1}}$ \\
		$K_\text{d,Casr}^\text{eq}$ & 2.24mM & $k^+ = 1.2996\times 10^{11}\si{s^{-1}}$\\& & $k^- = 2.5900\times 10^{10}\si{mM^{-2} s^{-1}}$ \\
		$K_\text{d,H1}^\text{eq}$ & $1.09 \times 10^{-5} \si{mM}$& $k^+ = 2.3761\times 10^{15}\si{mM^{-1} s^{-1}}$ \\& & $k^- = 2.5900\times 10^{10}\si{s^{-1}}$ \\
		$K_\text{d,Hi}^\text{eq}$ & $3.54 \times 10^{-3} \si{mM^2}$ & $k^+ =2.5900\times 10^{10}\si{s^{-1}}$\\& & $k^- = 7.3164\times 10^{12}\si{mM^{-2} s^{-1}}$ \\
		$K_\text{d,Hsr}^\text{eq}$ & $1.05 \times 10^{-8} \si{mM^2}$& $k^+ = 2.4667\times 10^{18}\si{mM^{-2} s^{-1}}$ \\& & $k^- = 2.5900\times 10^{10}\si{s^{-1}}$ \\
		$K_\text{d,H}^\text{eq}$ & $7.24 \times 10^{-5} \si{mM}$ & $k^+ =2.5900\times 10^{10}\si{s^{-1}}$\\& & $k^- = 3.5773\times 10^{14}\si{mM^{-1} s^{-1}}$ \\
		$n$ & 2.0 & - \\ \bottomrule
	\end{tabular}
	\label{tab:Tran_parameters}
\end{table}

\begin{table}[H]
	\caption{\textbf{Parameters for the bond graph model of the SERCA pump.} The reaction parameters have unit $\si{fmol/s}$, and the species parameters have unit $\si{fmol^{-1}}$.}
	\centering
	\begin{tabular}{l l l}
		\toprule
		\textbf{Species/Reaction} & \textbf{Parameter} & \textbf{Value} \\ \midrule
		$\text{R}1 \rightarrow 2$&  $\kappa_{12}$ & $0.00053004$\\ 
		$\text{R}2 \rightarrow 4$&$\kappa_{24}$ & $1567.7476 $\\ 
		$\text{R}2 \rightarrow 2a$&$\kappa_\mathrm{22a}$ & $8326784.0537 $\\ 
		$\text{R}4 \rightarrow 5$&$\kappa_{45}$ & $1567.7476$\\ 
		$\text{R}5 \rightarrow 6$&$\kappa_{56}$ & $3063.4006 $\\ 
		$\text{R}6 \rightarrow 8$&$\kappa_{68}$ & $130852.3839 $\\ 
		$\text{R}8 \rightarrow 9$&$\kappa_{89}$ & $11612934.8748 $\\ 
		$\text{R}9 \rightarrow 10$&$\kappa_{910}$ & $11612934.8748 $\\ 
		$\text{R}10 \rightarrow 1$&$\kappa_{101}$ & $0.049926 $\\ 
		$\text{P}_1$ &$K_1$ & $5263.6085 $\\ 
		$\text{P}_2$ &$K_2$ & $3803.6518 $\\ 
		$\text{P}_\mathrm{2a}$ &$K_\mathrm{2a}$ & $3110.4445$ \\ 
		$\text{P}_4$ &$K_4$ & $16520516.1239$ \\ 
		$\text{P}_5$ &$K_5$ & $0.82914 $\\ 
		$\text{P}_6$ &$K_6$ & $993148.433 $\\ 
		$\text{P}_8$ &$K_8$ & $37.7379 $\\ 
		$\text{P}_9$ &$K_9$ & $2230.2717 $\\ 
		$\text{P}_{10}$ &$K_{10}$ & $410.6048$ \\ 
		$\text{H}^+$ &$K_H$ & $1862.5406$ \\ 
		$\text{Ca}_\text{i}$ &$K_\text{Cai}$ & $1.9058 $\\ 
		$\text{Ca}_\text{sr}$ &$K_\text{Casr}$ & $31.764$ \\ 
		$\text{MgATP}$ &$K_\text{MgATP}$ & $244.3021 $\\ 
		$\text{MgADP}$ &$K_\text{MgADP}$ & $5.8126 \times 10^{-7}$ \\ 
		$\text{P}_\text{i}$ &$K_\text{Pi}$ & $0.014921$ \\  \bottomrule
	\end{tabular}
	\label{tab:bg_parameters_SERCA}
\end{table}

\setcounter{table}{0}
\section{Stoichiometry and parameters for Na\tsp{}/K\tsp{} ATPase}
\label{sec:NaK_params}

\begin{table}[H]
	\centering
	\caption{\textbf{Forward stoichiometric matrix for the \citet{terkildsen_balance_2007} model of Na\tsp{}/K\tsp{} ATPase.}}
	\label{tab:NaK_Nf}
	\begin{tabular}{cccccccccccccccc}
		\toprule
		& \multicolumn{15}{c}{Reactions} \\ \cline{2-16}
		& 1   & 2 & 3 & 4 & 5 & 6 & 7 & 8 & 9 & 10 & 11 & 12 & 13 & 14 & 15     \\ \midrule
		P1    & 1 & 0 & 0 & 0 & 0 & 0 & 0 & 0 & 0 & 0 & 0 & 0 & 0 & 0 & 0 \\
		P2    & 0 & 1 & 0 & 0 & 0 & 0 & 0 & 0 & 0 & 0 & 0 & 0 & 0 & 0 & 0 \\
		P3    & 0 & 0 & 1 & 0 & 0 & 0 & 0 & 0 & 0 & 0 & 0 & 0 & 0 & 0 & 0 \\
		P4    & 0 & 0 & 0 & 1 & 0 & 0 & 0 & 0 & 0 & 0 & 0 & 0 & 0 & 0 & 0 \\
		P5    & 0 & 0 & 0 & 0 & 1 & 0 & 0 & 0 & 0 & 0 & 0 & 0 & 0 & 0 & 0 \\
		P6    & 0 & 0 & 0 & 0 & 0 & 1 & 0 & 0 & 0 & 0 & 0 & 0 & 0 & 0 & 0 \\
		P7    & 0 & 0 & 0 & 0 & 0 & 0 & 1 & 0 & 0 & 0 & 0 & 0 & 0 & 0 & 0 \\
		P8    & 0 & 0 & 0 & 0 & 0 & 0 & 0 & 1 & 0 & 0 & 0 & 0 & 0 & 0 & 0 \\
		P9    & 0 & 0 & 0 & 0 & 0 & 0 & 0 & 0 & 1 & 0 & 0 & 0 & 0 & 0 & 0 \\
		P10   & 0 & 0 & 0 & 0 & 0 & 0 & 0 & 0 & 0 & 1 & 0 & 0 & 0 & 0 & 0 \\
		P11   & 0 & 0 & 0 & 0 & 0 & 0 & 0 & 0 & 0 & 0 & 1 & 0 & 0 & 0 & 0 \\
		P12   & 0 & 0 & 0 & 0 & 0 & 0 & 0 & 0 & 0 & 0 & 0 & 1 & 0 & 0 & 0 \\
		P13   & 0 & 0 & 0 & 0 & 0 & 0 & 0 & 0 & 0 & 0 & 0 & 0 & 1 & 0 & 0 \\
		P14   & 0 & 0 & 0 & 0 & 0 & 0 & 0 & 0 & 0 & 0 & 0 & 0 & 0 & 1 & 0 \\
		P15   & 0 & 0 & 0 & 0 & 0 & 0 & 0 & 0 & 0 & 0 & 0 & 0 & 0 & 0 & 1 \\
		$\mathrm{K^+_i}$  & 0 & 0 & 0 & 0 & 0 & 0 & 0 & 0 & 0 & 0 & 0 & 0 & 0 & 0 & 0 \\
		$\mathrm{K^+_e}$   & 0 & 0 & 0 & 0 & 0 & 0 & 0 & 0 & 0 & 0 & 1 & 1 & 0 & 0 & 0 \\
		$\mathrm{Na^+_i}$  & 0 & 0 & 1 & 1 & 1 & 0 & 0 & 0 & 0 & 0 & 0 & 0 & 0 & 0 & 0 \\
		$\mathrm{Na^+_e}$  & 0 & 0 & 0 & 0 & 0 & 0 & 0 & 0 & 0 & 0 & 0 & 0 & 0 & 0 & 0 \\
		MgATP & 0 & 0 & 0 & 0 & 0 & 0 & 0 & 0 & 0 & 0 & 0 & 0 & 0 & 1 & 0 \\
		MgADP & 0 & 0 & 0 & 0 & 0 & 0 & 0 & 0 & 0 & 0 & 0 & 0 & 0 & 0 & 0 \\
		Pi    & 0 & 0 & 0 & 0 & 0 & 0 & 0 & 0 & 0 & 0 & 0 & 0 & 0 & 0 & 0 \\
		H\tsp{}     & 0 & 0 & 0 & 0 & 0 & 0 & 0 & 0 & 0 & 0 & 0 & 0 & 0 & 0 & 0 \\ \bottomrule
	\end{tabular}
\end{table}

\begin{table}[H]
	\centering
	\caption{\textbf{Reverse stoichiometric matrix for the \citet{terkildsen_balance_2007} model of Na\tsp{}/K\tsp{} ATPase.}}
	\label{tab:NaK_Nr}
	\begin{tabular}{cccccccccccccccc}
		\toprule
		& \multicolumn{15}{c}{Reactions} \\ \cline{2-16}
		& 1   & 2 & 3 & 4 & 5 & 6 & 7 & 8 & 9 & 10 & 11 & 12 & 13 & 14 & 15     \\ \midrule
		P1    & 0   & 0   & 0   & 0   & 0   & 0   & 0   & 0   & 0    & 0    & 0    & 0    & 0    & 0    & 1 \\
		P2    & 1   & 0   & 0   & 0   & 0   & 0   & 0   & 0   & 0    & 0    & 0    & 0    & 0    & 0    & 0 \\
		P3    & 0   & 1   & 0   & 0   & 0   & 0   & 0   & 0   & 0    & 0    & 0    & 0    & 0    & 0    & 0 \\
		P4    & 0   & 0   & 1   & 0   & 0   & 0   & 0   & 0   & 0    & 0    & 0    & 0    & 0    & 0    & 0 \\
		P5    & 0   & 0   & 0   & 1   & 0   & 0   & 0   & 0   & 0    & 0    & 0    & 0    & 0    & 0    & 0 \\
		P6    & 0   & 0   & 0   & 0   & 1   & 0   & 0   & 0   & 0    & 0    & 0    & 0    & 0    & 0    & 0 \\
		P7    & 0   & 0   & 0   & 0   & 0   & 1   & 0   & 0   & 0    & 0    & 0    & 0    & 0    & 0    & 0 \\
		P8    & 0   & 0   & 0   & 0   & 0   & 0   & 1   & 0   & 0    & 0    & 0    & 0    & 0    & 0    & 0 \\
		P9    & 0   & 0   & 0   & 0   & 0   & 0   & 0   & 1   & 0    & 0    & 0    & 0    & 0    & 0    & 0 \\
		P10   & 0   & 0   & 0   & 0   & 0   & 0   & 0   & 0   & 1    & 0    & 0    & 0    & 0    & 0    & 0 \\
		P11   & 0   & 0   & 0   & 0   & 0   & 0   & 0   & 0   & 0    & 1    & 0    & 0    & 0    & 0    & 0 \\
		P12   & 0   & 0   & 0   & 0   & 0   & 0   & 0   & 0   & 0    & 0    & 1    & 0    & 0    & 0    & 0 \\
		P13   & 0   & 0   & 0   & 0   & 0   & 0   & 0   & 0   & 0    & 0    & 0    & 1    & 0    & 0    & 0 \\
		P14   & 0   & 0   & 0   & 0   & 0   & 0   & 0   & 0   & 0    & 0    & 0    & 0    & 1    & 0    & 0 \\
		P15   & 0   & 0   & 0   & 0   & 0   & 0   & 0   & 0   & 0    & 0    & 0    & 0    & 0    & 1    & 0 \\
		$\mathrm{K^+_i}$  & 1   & 1   & 0   & 0   & 0   & 0   & 0   & 0   & 0    & 0    & 0    & 0    & 0    & 0    & 0 \\
		$\mathrm{K^+_e}$  & 0   & 0   & 0   & 0   & 0   & 0   & 0   & 0   & 0    & 0    & 0    & 0    & 0    & 0    & 0 \\
		$\mathrm{Na^+_i}$ & 0   & 0   & 0   & 0   & 0   & 0   & 0   & 0   & 0    & 0    & 0    & 0    & 0    & 0    & 0 \\
		$\mathrm{Na^+_e}$ & 0   & 0   & 0   & 0   & 0   & 0   & 0   & 1   & 1    & 1    & 0    & 0    & 0    & 0    & 0 \\
		MgATP & 0   & 0   & 0   & 0   & 0   & 0   & 0   & 0   & 0    & 0    & 0    & 0    & 0    & 0    & 0 \\
		MgADP & 0   & 0   & 0   & 0   & 0   & 1   & 0   & 0   & 0    & 0    & 0    & 0    & 0    & 0    & 0 \\
		Pi    & 0   & 0   & 0   & 0   & 0   & 0   & 0   & 0   & 0    & 0    & 0    & 0    & 1    & 0    & 0 \\
		H\tsp{}     & 0   & 0   & 0   & 0   & 0   & 0   & 0   & 0   & 0    & 0    & 0    & 0    & 1    & 0    & 0 \\ \bottomrule
	\end{tabular}
\end{table}

\begin{table}[H]
	\caption{\textbf{Kinetic parameters for the updated cardiac Na$^+$/K$^+$ ATPase model.}}
	\centering
	\bgroup
	\def\arraystretch{1.3}
	\begin{tabular}{c p{0.48\linewidth} l}
		\toprule
		\textbf{Parameter} & \textbf{Description}& \textbf{Value} \\ \midrule
		$k_1^+$ & Forward rate constant of reaction R6 & $1423.2\ \si{s^{-1}}$\\ 
		$k_1^-$ & Reverse rate constant of reaction R6 & $225.9048\ \si{s^{-1}}$\\ 
		$k_2^+$ & Forward rate constant of reaction R7 & $11564.8064\ \si{s^{-1}}$\\ 
		$k_2^-$ & Reverse rate constant of reaction R7 & $36355.3201\ \si{s^{-1}}$\\ 
		$k_3^+$ & Forward rate constant of reaction R13 & $194.4506\ \si{s^{-1}}$\\ 
		$k_3^-$ & Reverse rate constant of reaction R13 & $281037.2758\ \si{mM^{-2}s^{-1}}$\\ 
		$k_4^+$ & Forward rate constant of reaction R15 & $30629.8836\ \si{s^{-1}}$\\ 
		$k_4^-$ & Reverse rate constant of reaction R15 & $1.574\times 10^{6}\ \si{s^{-1}}$\\ 
		$K_{\text{d,Nai}}^0$ & Voltage-dependent dissociation constant of intracellular $\mathrm{Na^+}$  & $579.7295\ \si{mM}$\\ 
		$K_{\text{d,Nae}}^0$ & Voltage-dependent dissociation constant of extracellular $\mathrm{Na^+}$ & $0.034879\ \si{mM}$\\ 
		$K_{\text{d,Nai}}$ & Voltage-independent dissociation constant of intracellular $\mathrm{Na^+}$ & $5.6399\ \si{mM}$\\ 
		$K_{\text{d,Nae}}$ & Voltage-independent dissociation constant of extracellular $\mathrm{Na^+}$  & $10616.9377\ \si{mM}$\\ 
		$K_{\text{d,Ki}}$ & Dissociation constant of intracellular $\mathrm{K^+}$ & $16794.976\ \si{mM}$\\ 
		$K_{\text{d,Ke}}$ & Dissociation constant of extracellular $\mathrm{K^+}$ & $1.0817\ \si{mM}$\\ 
		$K_{\text{d,MgATP}}$ & Dissociation constant of MgATP & $140.3709\ \si{mM}$\\ 
		$\Delta$ & Charge translocated by reaction R5 & $-0.0550$ \\
		Pump density & Number of pumps per $\si{\mu m^2}$  & $1360.2624\ \si{\mu m^{-2}}$  \\ \bottomrule& 
	\end{tabular}
	\egroup
	\label{tab:Terkildsen_parameters}
\end{table}

\begin{table}[H]
	\caption{\textbf{Parameters for the bond graph version of the updated cardiac Na$^+$/K$^+$ ATPase model.}}
	\centering
	\small
	\begin{tabular}{cl c l}
		\toprule
		\textbf{Component} & \textbf{Description}& \textbf{Parameter} & \textbf{Value} \\ \midrule
		R1 & Reaction R1 & $\kappa_1$ & $ 330.5462\ \si{fmol/s} $\\ 
		R2 & Reaction R2 & $\kappa_2$ & $ 132850.9145\ \si{fmol/s} $\\ 
		R3 & Reaction R3 & $\kappa_3$ & $ 200356.0223\ \si{fmol/s} $\\ 
		R4 & Reaction R4 & $\kappa_4$ & $ 2238785.3951\ \si{fmol/s} $\\ 
		R5 & Reaction R5 & $\kappa_5$ & $ 10787.9052\ \si{fmol/s} $\\ 
		R6 & Reaction R6 & $\kappa_6$ & $ 15.3533\ \si{fmol/s} $\\ 
		R7 & Reaction R7 & $\kappa_7$ & $ 2.3822\ \si{fmol/s} $\\ 
		R8 & Reaction R8 & $\kappa_8$ & $ 2.2855\ \si{fmol/s} $\\ 
		R9 & Reaction R9 & $\kappa_9$ & $ 1540.1349\ \si{fmol/s} $\\ 
		R10 & Reaction R10 & $\kappa_{10}$ & $ 259461.6507\ \si{fmol/s} $\\ 
		R11 & Reaction R11 & $\kappa_{11}$ & $ 172042.3334\ \si{fmol/s} $\\ 
		R12 & Reaction R12 & $\kappa_{12}$ & $ 6646440.3909\ \si{fmol/s} $\\ 
		R13 & Reaction R13 & $\kappa_{13}$ & $ 597.4136\ \si{fmol/s} $\\ 
		R14 & Reaction R14 & $\kappa_{14}$ & $ 70.9823\ \si{fmol/s} $\\ 
		R15 & Reaction R15 & $\kappa_{15}$ & $ 0.015489\ \si{fmol/s} $\\ 
		$\text{P}_1$ & Pump state ATP--E\textsubscript{i} K\textsubscript{2}
		& $K_1$ & $101619537.2009\ \si{fmol^{-1}}$ \\ 
		$\text{P}_2$ & Pump state ATP--E\textsubscript{i} K\textsubscript{1}
		& $K_2$ & $63209.8623\ \si{fmol^{-1}}$ \\ 
		$\text{P}_3$ & Pump state ATP--E\textsubscript{i}
		& $K_3$ & $157.2724\ \si{fmol^{-1}}$ \\ 
		$\text{P}_4$ & Pump state ATP--E\textsubscript{i} Na\textsubscript{1}
		& $K_4$ & $14.0748\ \si{fmol^{-1}}$ \\ 
		$\text{P}_5$ & Pump state ATP--E\textsubscript{i} Na\textsubscript{2}
		& $K_5$ & $5.0384\ \si{fmol^{-1}}$ \\ 
		$\text{P}_6$ & Pump state ATP--E\textsubscript{i} Na\textsubscript{3}
		& $K_6$ & $92.6964\ \si{fmol^{-1}}$ \\ 
		$\text{P}_7$ & Pump state P--E\textsubscript{i} (Na\textsubscript{3})
		& $K_7$ & $4854.5924\ \si{fmol^{-1}}$ \\ 
		$\text{P}_8$ & Pump state P--E\textsubscript{e} Na\textsubscript{3}
		& $K_8$ & $15260.9786\ \si{fmol^{-1}}$ \\ 
		$\text{P}_9$ & Pump state P--E\textsubscript{e} Na\textsubscript{2}
		& $K_9$ & $13787022.8009\ \si{fmol^{-1}}$ \\ 
		$\text{P}_{10}$ & Pump state P--E\textsubscript{e} Na\textsubscript{1}
		& $K_{10}$ & $20459.5509\ \si{fmol^{-1}}$ \\ 
		$\text{P}_{11}$ & Pump state P--E\textsubscript{e}
		& $K_{11}$ & $121.4456\ \si{fmol^{-1}}$ \\ 
		$\text{P}_{12}$ & Pump state P--E\textsubscript{e} K\textsubscript{1}
		& $K_{12}$ & $3.1436\ \si{fmol^{-1}}$ \\ 
		$\text{P}_{13}$ & Pump state P--E\textsubscript{e} K\textsubscript{2}
		& $K_{13}$ & $0.32549\ \si{fmol^{-1}}$ \\ 
		$\text{P}_{14}$ & Pump state E\textsubscript{e} (K\textsubscript{2})
		& $K_{14}$ & $156.3283\ \si{fmol^{-1}}$ \\ 
		$\text{P}_{15}$ & Pump state ATP--Ee (K\textsubscript{2})
		& $K_{15}$ & $1977546.8577\ \si{fmol^{-1}}$ \\ 
		Ki & Intracellular $\text{K}_\text{i}^+$ & $K_\text{Ki}$ & $0.0012595\ \si{fmol^{-1}}$ \\ 
		Ke & Extracellular $\text{K}_\text{e}^+$ & $K_\text{Ke}$ & $0.009236\ \si{fmol^{-1}}$ \\ 
		Nai & Intracellular $\text{Na}_\text{i}^+$ & $K_\text{Nai}$ & $0.00083514\ \si{fmol^{-1}}$ \\ 
		Nae & Extracellular $\text{Na}_\text{e}^+$ & $K_\text{Nae}$ & $0.0061242\ \si{fmol^{-1}}$ \\ 
		$\text{MgATP}$ & Intracellular MgATP & $K_\text{MgATP}$ & $2.3715\ \si{fmol^{-1}}$ \\ 
		$\text{MgADP}$ & Intracellular MgADP & $K_\text{MgADP}$ & $7.976 \times 10^{-5} \ \si{fmol^{-1}}$ \\ 
		$\text{P}_\text{i}$ & Free inorganic phosphate & $K_\text{Pi}$ & $0.04565\ \si{fmol^{-1}}$ \\ 
		H & Intracellular $\text{H}^+$ & $K_\text{H}$ & $0.04565\ \si{fmol^{-1}}$ \\ 
		mem & Membrane capacitance & $C_m$ & $153400\ \si{fF}$ \\
		zF{\_}5 & Charge translocated by R5 & $z_5$ & $-0.0550$ \\
		zF{\_}8 & Charge translocated by R8 & $z_8$ & $-0.9450$ \\ \bottomrule& & 
	\end{tabular}
	\label{tab:bg_parameters_NaK}
\end{table}

\end{document}